\documentclass[a4paper,11pt]{article}
\usepackage[table,xcdraw]{xcolor}
\usepackage{jcappub}
\usepackage[utf8]{inputenc}
\usepackage{amsmath, amssymb, mathtools}
\usepackage{tensor}
\usepackage{graphicx}
\usepackage{subcaption}
\usepackage{epsfig}
\usepackage{adjustbox}
\usepackage{tikz}
\usepackage{physics}
\usepackage{lipsum}  
\usepackage{bigints}
\usepackage{verbatim}
\usepackage{bbold}
\usepackage{color}
\usepackage{multicol}
\usepackage{array}
\usepackage{boldline}
\usepackage{makecell}
\usepackage{multirow}
\usepackage{booktabs}  % for \hlineB
\usepackage{xparse}    % if you define \leri yourself
\usepackage{colortbl}

\newcolumntype{P}[1]{>{\raggedright\arraybackslash}p{#1}}
\newcolumntype{C}[1]{>{\centering\arraybackslash}p{#1}}
\usepackage{booktabs}
\newcolumntype{V}[1]{!{\vrule width #1pt}}

\usepackage[normalem]{ulem}
\usepackage{caption}

\usepackage{subcaption}
\newcommand{\leri}[1]{\left(#1\right)}
\newcommand{\lerisq}[1]{\left[#1\right]}

\newcommand{\jord}{\frac{1}{2\kappa^2}\int d^4x\,\sqrt{-g}\;}
\newcommand{\ein}{\frac{1}{2\kappa^2}\int d^4x\,\sqrt{-q}\;}
\newcommand{\pal}{\mathcal{R}}

\title{Dynamical system analysis of the cosmological phases in Palatini $k$-essence gravity}

\author[a]{F. Moretti,\note{Corresponding author.}}
\author[b,1]{F. Bombacigno}

\affiliation[a]{Fusion and Nuclear Safety Department, ENEA, C. R. Frascati,
\\Via E. Fermi 45, 00044 Frascati, Italy}
\affiliation[b]{Instituto de Física Corpuscular (IFIC), CSIC-Universitat de València,\\C/ Doctor Moliner 50, 46100, Burjassot, Spain}

\emailAdd{fabio.moretti.1@enea.it}
\emailAdd{flavio2.bombacigno@uv.es}

\abstract{We formulate a generalized $k$-essence model in the presence of a Palatini $f(\pal)$ gravitational sector. In the corresponding biscalar-tensor theory, we discuss the distinguished dynamical properties of the two scalar fields, elucidating how the "Palatini" scalaron can be still algebraically solved in terms of matter, the $k$-essence field and its kinetic term. We derive the conditions ensuring the absence of Ostrogradsky modes and the well-posedness of the initial data problem, also providing an intriguing analogy with a specific class of DHOST theories. Then, we investigate the cosmology of a flat Friedmann-Lemaître–Robertson-Walker spacetime according to a dynamical system approach, with the aim of determining the set of fixed points in the phase space, representing specific periods of the Universe evolution and characterized by different effective barotropic index $w_{\text{eff}}$. The analysis reveals the presence of a range of possible configurations, with the existence of (quasi) de-Sitter epochs connected by heteroclinic orbits, scaling solutions and quintessence phases.}

\makeatletter
\gdef\@fpheader{}
\makeatother
\begin{document}
\maketitle
\flushbottom
\section{Introduction}
Dynamical systems techniques encompass a variety of versatile mathematical tools, able to provide a qualitative description of the evolution of the states of the system, as they are spanned across the phase space. These methods allow for the characterization of the critical points of the system and their stability, together with the identification of possible trajectories connecting the different available configurations~\cite{wiggins1990,roberts2015model}. When applied to cosmology~\cite{Bahamonde:2017ize}, this approach facilitates the depiction of the Universe in its entirety, in terms of the dynamical behaviour of characteristic variables, possibly built out of fundamental physical quantities (see the discussion in \autoref{sec: 4}). In order for this representation to be consistent, a mathematical law governing the evolution of the states must be assigned, as it is provided for instance by the Einstein equations in the context of General Relativity (GR). Even if they currently constitute our best theoretical setting for describing gravitational interactions, observations have been mounting up at over the last years of phenomena at cosmological scales still lacking for a satisfactory description. For the purposes of this work, we refer in particular to the present accelerated expansion of the Universe~\cite{SupernovaCosmologyProject:1998vns,SupernovaSearchTeam:1998fmf}, reconciled within the theoretical framework of GR by the introduction of a cosmological constant term. This is not clear of conceptual limitations~\cite{Sahni:1999gb, Carroll:2000fy, Peebles:2002gy, Padmanabhan:2002ji, Copeland:2006wr, Caldwell:2009ix, Li:2011sd, Martin:2012bt, Weinberg:1988cp, Krauss:1995yb, Weinberg:2000yb, Sahni:2002kh, Yokoyama:2003ii, Nobbenhuis:2004wn, Burgess:2013ara, Joyce:2014kja, Bull:2015stt, Wang:2016lxa, Brustein:1992nk, Witten:2000zk, Kachru:2003aw, Polchinski:2006gy, Danielsson:2018ztv, Zlatev:1998tr, Pavon:2005yx, Velten:2014nra}, and it has been recently challenged by the measurements of the DESI collaboration~\cite{DESI:2024mwx,Giare:2024oil,Giare:2024gpk,DESI:2025fii,DESI:2025wyn,DESI:2025zgx}, with evidence for a dynamical dark energy setting, where the state parameter $w$ crosses the phantom divide line $w = -1$.
\\In the aim of explaining such an accelerated expansion of the Universe without resorting to dark energy, several modifications to GR have been formulated, and a dynamical system analysis has been widely adopted for describing these alternative cosmological phase spaces. These appear for instance\footnote{This list represents a selection of works directly related to dynamical system approaches in $f(R)$ gravity and its generalizations. For a comprehensive review on the existing literature, we remind the reader to \cite{Bahamonde:2017ize}.} in $f(R)$ gravity (in its metric and Palatini formulation)~\cite{Amendola:2006we,Amendola:2007nt,deSouza:2007zpn,Fay:2007gg,Alho:2016gzi,Odintsov:2017tbc,Chakraborty:2018bxh,Chakraborty:2018ost}, along with its extension involving non-minimal geometry-matter coupling~\cite{Shabani:2013djy,Ribeiro:2014sla,Azevedo:2016ehy}, hybrid metric-Palatini formulation~\cite{Carloni:2015bua,Rosa:2017jld,Rosa:2019ejh}, (symmetric) teleparallel applications \cite{Otalora:2013dsa,Skugoreva:2014ena,Carloni:2015lsa,Khyllep:2022spx,Duchaniya:2022fmc,Csillag:2025gnz,Murtaza:2025gme}, higher-order corrections~\cite{Carroll:2004de,Alimohammadi:2009js,Chatzarakis:2019fbn}, and non-canonical kinetic terms for additional scalar fields~\cite{Oikonomou:2019muq}. In this work we deal with this last class of theories, commonly known as $k$-essence gravity ~\cite{Nojiri:2019dqc,Oikonomou:2019muq}, where the Lagrangian density of GR is promoted to a generic function $f(R,\xi,X)$ of the Ricci scalar $R$, a scalar field $\xi$ and its kinetic term $X=g^{\mu\nu}\nabla_\mu\xi\nabla_\nu\xi$. 
These models have also proved to exhibit viable inflationary attractors, reproducing early (quasi)de-Sitter epochs, which can become asymptotically unstable and provide reliable mechanisms for a graceful exit from inflation. 
\\The studies in~\cite{Nojiri:2019dqc,Oikonomou:2019muq} rely on a metric formulation~\cite{Olmo:2005hc}, where the metric field $g_{\mu\nu}$ is the only gravitational degree of freedom, besides the scalar field, and the affine connection is assumed to be the Levi-Civita one. A different perspective, also known as Palatini formulation, allows for a different geometric structure, where the affine connection is an independent dynamical entity with respect to the metric field. Here the connection enters the definition of Riemann-derived quantities and it does not directly couple to matter fields\footnote{In the following we adopt the standard notation $R$ for denoting purely metric quantities, and the calligraphic font $\pal$ for geometric objects in the Palatini formalism.}. In the context of pure $f(R)$ gravity the two different approaches have been demonstrated to produced quite different dynamical settings, characterized by an additional gravitational scalar degree of freedom or additional source terms in the metric and Palatini case, respectively~\cite{Sotiriou2007,Olmo:2011uz}. In the latter case, in particular, a solution for the connection can be still obtained, resulting in the Levi-Civita connection of an auxiliary metric, related to the original metric by a conformal transformation. The Palatini formulation of $k$-essence $f(\pal,\xi,X)$ gravity has been poorly addressed, and preliminary steps in this direction have been moved in~\cite{Galtsov:2018xuc,Kubota:2020ehu,Dioguardi:2023jwa,Dimopoulos:2025fuq} for very specific configurations. 
\\ In the first part of this work we consider a generic $f(\pal,\xi,X)$ model, that we conveniently rewrite in a scalar-tensor form, by introducing the generalized potential term $U(\phi,\xi,X)$, where $\phi$ denotes the typical Palatini scalaron of the Jordan frame, non-minimally coupled to the Ricci scalar. We address the problem of the stability of the theory, and we derive the set of conditions for the function $U$ ensuring the absence of pathological ghost modes and a well defined dynamical evolution. In vacuum, we also establish an interesting correspondence of our model with some class of degenerate higher order scalar-tensor theories (DHOST) \cite{BenAchour:2016cay,Langlois:2015cwa}, compatible with the phenomenological constraints from multimessenger observations \cite{Langlois:2017mxy,Langlois:2017dyl}.
\\In the second part, we restrict our attention to specific realizations of the theory, where the structural equation does not depend on the kinetic term $X$, but the Palatini scalar is still non-minimally coupled with $\xi$ in the Lagrangian. For such a configuration, we perform a dynamical system analysis of the resulting cosmological phase space. In particular, we consider two distinguished scenarios, characterized by a two- or single-component perfect fluid, both for a power-law and a exponential coupling between the $k$-essence field and the Palatini scalar. We find that this model offers a rich and complex phenomenology when applied to cosmology. The equilibrium points investigated can be subdivided into three main categories: $i)$ de Sitter attractors/saddles, in which the dynamics is dominated by a constant scalar field $\xi$ and the scale factor has an exponential behavior; $ii) $ scaling solutions, where one of the components of the cosmological fluid is the main driver of the cosmological expansion, reproducing matter and radiation dominated eras; $iii)$ equilibria in which the kinetic and potential energy of $\xi$ are dominant: here the effective barotropic index, describing the physics associated with the cosmological dynamics, can be tuned via the theory parameter, making these solutions potential mimickers of fluid-dominated phases. The great variety of features that the Palatini formulation of $k$-essence $f\leri{\mathcal{R},\xi,X}$ gravity offers in the cosmological context makes this model particularly interesting, for its ability in reproducing different phases of the Universe history, and, as we will show in the subsequent sections, an intriguing link between inflationary and late-time de Sitter expansions.

The manuscript is organized as follows. In \autoref{sec: 2} we introduce the model, outlining the main properties and deriving the equations of motion for the different fields; in \autoref{sec: 3} we derive the conditions on the function $U$ enforcing the absence of ghost modes and a well-defined initial data problem, also, we highlight the relation with DHOST models; in \autoref{sec: 4} we specialize the discussion to a specific configuration of $U$, and we reformulate the problem in the formalism of the dynamical systems; in \autoref{sec: 5} and \autoref{sec: 6} we consider the cases of an energy momentum tensor with two and one fluid, respectively; in \autoref{sec: 7} we reconstruct the original Lagrangian density corresponding to the forms of the potential discussed; eventually, in \autoref{sec: 8} we discuss the results and comment some possible future lines of investigation.
\\The Riemann curvature tensor is defined in terms of the independent connection as $\mathcal{R}\indices{^\rho_{\mu\sigma\nu}}=\partial_\sigma\Gamma\indices{^\rho_{\mu\nu}}-\partial_\nu\Gamma\indices{^\rho_{\mu\sigma}}+\Gamma\indices{^\rho_{\tau\sigma}}\Gamma\indices{^\tau_{\mu\nu}}-\Gamma\indices{^\rho_{\tau\nu}}\Gamma\indices{^\tau_{\mu\sigma}}$. Spacetime signature is chosen mostly plus ($\eta_{\mu\nu}=\text{diag}(-1,1,1,1)$) and the gravitational coupling set as $\kappa^2=8\pi G$, with $c=1$.

\section{The generalized Palatini $k$-essence model}\label{sec: 2}
\noindent 
The starting point of our analysis is a Palatini formulation of the $k$-essence action introduced in~\cite{Oikonomou:2019muq}, i.e.
\begin{equation}
    S=\jord f(\pal,\xi,X),
    \label{eq: action f(R)}
\end{equation}
where the Palatini Ricci curvature $\mathcal{R}$ is assumed to depend on the independent connection $\Gamma\indices{^\rho_{\mu\nu}}$ as
\begin{align}
    \pal=g^{\mu\nu}\pal_{\mu\nu}(\Gamma),
\end{align}
with $\pal_{\mu\nu}$ denoting the trace of the Riemann tensors on the first and third indices. The kinetic term for the scalar field $\xi$ is displayed by $X\equiv g^{\mu\nu}\nabla_\mu\xi\nabla_\nu\xi$. A convenient scalar-tensor reformulation of \eqref{eq: action f(R)} can be easily obtained by following the standard procedure for ordinary $f(R)$ theories (see \cite{Sotiriou2007}), resulting in
\begin{equation}
    S=\jord \leri{\phi\pal - U(\phi,\xi,X)},
    \label{eq: action scaten pal}
\end{equation}
where the potential-like term $U$ is defined by
\begin{equation}
    U(\phi,\xi,X) = \phi \pal(\phi,\xi,X) - f(\pal(\phi,\xi,X),\xi,X),
\end{equation}
with $\pal$ formally inverted from the definition of the scalar field $\phi = \frac{\partial f}{\partial \pal}$.
It is possible to show that the equation for the connection $\Gamma\indices{^\rho_{\mu\nu}}$, as it follows from \eqref{eq: action scaten pal}, takes the form
\begin{equation}
    \nabla_\rho \leri{\sqrt{-g}\,\phi\, g^{\mu\nu}}=0,
\end{equation}
where we assumed the hypothesis\footnote{We remark that we are not compelled to assume a priori a torsionless and metric-compatible connection, in that a dynamical equivalent result can be still obtained by solving the original equation of the connection for the different components of torsion and non-metricity. In this case, once the projective symmetry is exploited for gauging out the spurious degrees of freedom of the affine connection, the non-Riemannian parts of the geometry can be completely solved in terms of the derivatives of the scalar field $\phi$, so that the final expression of the Palatini curvature is still displayed by \eqref{eq: sol pal} (see \cite{Olmo:2005hc, Olmo:2011uz, Iosifidis:2018zjj} for technical details).} of a vanishing torsion ($\Gamma\indices{^\rho_{\mu\nu}}-\Gamma\indices{^\rho_{\nu\mu}}=0$) and metric compatibility ($\nabla_\rho g_{\mu\nu}=0$). The solution of such an equation is displayed by the Levi-Civita connection for the conformal metric $h_{\mu\nu}=\phi g_{\mu\nu}$, i.e.
\begin{equation}
\Gamma\indices{^\rho_{\mu\nu}}=\frac{1}{2}h^{\rho\sigma}\leri{\partial_\mu h_{\nu\sigma}+\partial_\nu h_{\mu\sigma}-\partial_\sigma h_{\mu\nu}},
\end{equation}
which can be equivalently rewritten in terms of the original metric $g_{\mu\nu}$ and the scalar field $\phi$ as
\begin{align}
    \Gamma\indices{^\rho_{\mu\nu}}=&\frac{1}{2}g^{\rho\sigma}\leri{\partial_\mu g_{\nu\sigma}+\partial_\nu g_{\mu\sigma}-\partial_\sigma g_{\mu\nu}}+\frac{1}{2\phi}\leri{\delta\indices{^\rho_\nu}\partial_\mu\phi+\delta\indices{^\rho_\mu}\partial_\nu\phi-g_{\mu\nu}\partial^\rho\phi}.
    \label{eq: sol connection expl}
\end{align}
It is thus possible to expand the Palatini curvature $\pal$ in terms of the metric Ricci scalar $R$ and the derivatives of the scalar field $\phi$, i.e.
\begin{equation}
    \pal=R+\frac{3\nabla_\mu\phi\nabla^\mu\phi}{2\phi^2}-\frac{\Box \phi}{\phi},
    \label{eq: sol pal}
\end{equation}
which plugged back into \eqref{eq: action scaten pal} results in
\begin{equation}
    S=\jord \leri{\phi R + \frac{3}{2\phi}(\nabla\phi)^2-U(\phi,\xi,X)}.
    \label{eq: action scaten met}
\end{equation}
The equation for the metric field obtained from the variation of \eqref{eq: action scaten met} with respect to the inverse metric $g^{\mu\nu}$ is displayed by
\begin{equation}
 \phi G_{\mu\nu}-(\nabla_\mu\nabla_\nu-g_{\mu\nu}\Box)\phi+\frac{3}{2\phi}\nabla_\mu\phi\nabla_\nu\phi-U_X\nabla_\mu\xi\nabla_\nu\xi-\frac{1}{2}g_{\mu\nu}\leri{\frac{3}{2\phi}(\nabla\phi)^2-U}=\kappa^2 T_{\mu\nu},
\label{eq: metric}
\end{equation}
while the equations for the scalar fields $\phi$ and $\xi$ result respectively in
\begin{equation}
    R+\frac{3}{2\phi^2}(\nabla\phi)^2-\frac{3\Box \phi}{\phi}-U_\phi = 0,
    \label{eq: scalar phi}
\end{equation}
and
\begin{equation}
\Box\xi-\frac{U_\xi}{2U_X}+\frac{\nabla^\mu\xi}{U_X}\leri{U_{X\phi}\nabla_\mu\phi+U_{X\xi}\nabla_\mu\xi+U_{XX}\nabla_\mu X}=0
\label{eq: scalar xi}.
\end{equation}
Here $T_{\mu\nu}$ (and its trace $T$) denotes the energy momentum tensor obtained from the matter Lagrangian possibly included in \eqref{eq: action f(R)}, i.e.
\begin{equation}
    T_{\mu\nu} = -\frac{2}{\sqrt{-g}}\frac{\delta (\sqrt{-g}\mathcal{L}_m)}{\delta g^{\mu\nu}}.
\end{equation}
Now, by combining \eqref{eq: scalar phi} with the trace of \eqref{eq: metric} we can obtain the structural equation for the general Palatini $f(\pal)$ $k$-essence model, i.e.
\begin{equation}
    2U-\phi \,U_\phi -X\, U_X=\kappa^2 T, 
    \label{eq: structural pal}
\end{equation}
that allows in principle to algebraically solve for the field $\phi$ in terms of both the matter and the remaining dynamical field $\xi$, i.e.
\begin{equation}
    \phi=\phi(\xi,X,T).
    \label{eq: Structural}
\end{equation}
Looking at \eqref{eq: scalar xi}, this relation implies that the dynamics of $\xi$ is affected by matter via the function $U$ and its field derivatives, once the solution \eqref{eq: Structural} is taken into account. Moreover, in general we expect relation \eqref{eq: structural pal} to introduce additional nonlinearities in the equation of $\xi$ via the argument $X$ in \eqref{eq: Structural}. 
\\We observe that there exist a special class of functions $U(\phi,\xi,X)$ ultimately leading to solution of the form $\phi = \phi(\xi,T)$. These correspond to the case where a pure potential term $W(\phi,\xi)$ can be singled out in $U$, apart from a coupling between some functions of $\phi$, $\xi$ and $X$. We are referring, specifically, to:
\begin{align}
    &U(\phi,\xi,X)=U_1(\phi)U_2(\xi)U_3(X) + W(\phi,\xi),
\end{align}
which inserted into \eqref{eq: structural pal} gives rise to
\begin{equation}
    2W-\phi W_{\phi}+U_2\leri{2U_1 U_2-\phi U_{1,\phi} U_3 - X U_1 U_{3,X}} = \kappa^2 T.
    \label{eq: structural no X}
\end{equation}
In order the dependence on $X$ to disappear in \eqref{eq: structural no X}, we must require the term between parentheses to vanish, which occurs for:
\begin{align}\label{formau}
    & U_3(X)=\beta_3 X^{2-m}, \quad U_1(\phi) =\beta_1 \phi^m\quad \text{for}\;\; m \neq 2
\end{align}
The simplest scenario endowed with a non-trivial coupling between the $k$-essence term and the "Palatini" scalar field $\phi$ is achieved for $m=1$, when the functions $U_3$ and $U_1$ depend linearly on their respective arguments. In this case it is then possible to reabsorb the function $U_2(\xi)$ in the kinetic term, via the antiderivative redefinition $\Xi(\xi) = \int \sqrt{U_2(\xi)}\; d\xi$. For this configuration, we can simply set $U_2(\xi) = 1$ without loss of generality, so that the function $U$ can be rearranged in the form:
\begin{equation}
    U(\phi,\xi,X)=\lambda^2\phi X+W(\phi,\xi),
    \label{eq: U final}
\end{equation}
where we redefined the constant as $\lambda^2 = \beta_1\beta_3$, in order to comply with the no-ghost condition discussed in  \autoref{sec: 3}. The implications of the choice \eqref{eq: U final} for a homogeneous and isotropic cosmological spacetime will be discussed in  \autoref{sec: 4}, where we pursue a dynamical system analysis for the space of the solutions. 
\\In the next section, instead, we derive some general properties for an unspecified $U(\phi,\xi,X)$ term, focusing on the conditions which guarantee the absence of instabilities and a well-defined dynamical evolution.

\section{Dynamical stability in Palatini $f(\pal)$ $k$-essence theory}\label{sec: 3}
The aim of this section is to determine the conditions for the theory to be devoid of dynamical instabilities and the equation of the $k$-essence field to be hyperbolic. These requirements amount to demand the absence of Ostrogradsky modes, eventually resulting in an Hamiltonian unbounded from below, and a Lorentzian signature for the effective metric defining the principal part, i.e. the wave operator, of the equation for $\xi$. This last request, in particular, ensures the well-posedness of the initial data problem, and as discussed in \cite{Babichev:2007dw,Babichev:2016hys,Babichev:2018twg} for the original $k$-essence theory, plays also a role in determining the causal properties of the local acoustic cone for the field perturbations.
\\Our analysis follows the presentation in \cite{Babichev:2007dw}, which we enlarge to include the contribution of auxiliary field $\phi$. We notice, indeed, that due to the structural equation, derivatives of $\phi$ can be formally rewritten as
\begin{equation}
    \nabla_\mu \phi = \phi_\xi \nabla_\mu \xi + \phi_X \nabla_\mu X+\phi_T \nabla_\mu T.
    \label{eq: eff derivative of phi}
\end{equation}
Without loss of generality, we can set $T_{\mu\nu}=0$ and consider the vacuum case, since by inspection of \eqref{eq: eff derivative of phi} we immediately see that the term proportional to $\phi_T$ does not affect the principal part. Under this hypothesis, the equation for $\xi$ can be rearranged as
\begin{equation}
Q^{\mu\nu}\nabla_\mu\nabla_\nu\xi+\leri{U_{X\phi}\phi_\xi+U_{X\xi}}X-\frac{U_\xi}{2}=0,
\label{eq: effective klein gordon xi}
\end{equation}
where we introduced the effective metric
\begin{equation}
     Q^{\mu\nu}(\xi,\nabla\xi)= U_Xg^{\mu\nu}+2\leri{U_{X\phi}\phi_X+U_{XX}}\nabla^\mu\xi\nabla^\nu\xi.
\end{equation}
It follows that when $\nabla_\mu\xi\neq 0$, the metric $Q^{\mu\nu}$ defining the principal part of the equation of motion is not conformally related to the spacetime metric $g^{\mu\nu}$. Therefore, we expect the wavefronts associated to the $\xi$-perturbations to define a different (acoustic) local causal structure with respect to that one determined by the original metric $g_{\mu\nu}$. We observe that in comparison with the standard $k$-essence theory \cite{Babichev:2007dw}, in this case assuming a Lagrangian linearly depending on the kinetic term for the scalar field is not sufficient for recovering a conformal relation. In our model, indeed, we have the additional contribution $U_{X\phi}\phi_X$, which in general does not vanish for $U\propto X$. A detailed discussion of the causal structure for the Palatini $f(\pal,\xi,X)$ theory evades the purposes of the current work, and we focus in the rest of this section on the conditions assuring the hyperbolic nature of the equation for the scalar field $\xi$.
\\In order the dynamics of $\xi$ to be not degenerate, the term proportional to $g^{\mu\nu}$ in the definition of $Q^{\mu\nu}$ must be not vanishing, leading us to the trivial condition $U_X\neq 0$. Its sign can be then determined by looking at the Einstein frame of the original action, defined by the conformal metric $q_{\mu\nu}=\phi g_{\mu\nu}$. Here, it takes the simpler form
\begin{equation}
    S=\ein \leri{R_E-  \frac{U(\phi,\xi,\phi\Tilde{X})}{\phi^2}},
    \label{eq: action einstein frame}
\end{equation}
where $\Tilde{X}\equiv q^{\mu\nu}\nabla_\mu\xi\nabla_\mu\xi=\phi^{-1}X$. For the metric $q_{\mu\nu}$ the equations of motion are given by
\begin{equation}
    G_{\mu\nu}(q)=\frac{ U_{X}\nabla_\mu\xi\nabla_\nu\xi}{\phi}-q_{\mu\nu}\frac{U}{2\phi^2},
\end{equation}
where we used $U_{\Tilde{X}}=\phi U_X$. Since the metric fields $g_{\mu\nu}$ and $q_{\mu\nu}$ are conformally related, they share the same set of null vectors $n^\mu$, i.e. $q_{\mu\nu}n^\mu n^\nu=\phi\; g_{\mu\nu}n^\mu n^\nu=0$. Then, because violation of the the null energy condition $T_{\mu\nu}n^\mu n^\nu \ge 0$ would imply an Hamiltonian unbounded from below, the requirement of stability leads to
\begin{equation}
    \frac{U_X (n^\mu \nabla_\mu \xi)^2}{\phi}\ge 0.
\end{equation}
Assuming $\phi>0$ in order to have a positive effective gravitational coupling $G/\phi$, this relation is then satisfied when $U_X> 0$. 
\\Under this hypothesis, let us introduce the metric $\Tilde{Q}^{\mu\nu} = U_X^{-1} Q^{\mu\nu}$, and let us investigate when such a metric is Lorentzian, i.e. when the equation for $\xi$ is hyperbolic. We start by observing that $\Tilde{Q}^{\mu\nu}$ admits an inverse $\Tilde{Q}_{\mu\nu}^{-1}$ given by
\begin{equation}
    \Tilde{Q}_{\mu\nu}^{-1}=g_{\mu\nu}-\frac{K}{1+KX}\nabla_\mu\xi\nabla_\nu\xi,
\end{equation}
where we defined $K\equiv \frac{2(U_{X\phi}\phi_X+U_{XX})}{U_X}$. For the inverse to be well-defined we must require $1+KX\neq 0$, which can be restated as
\begin{equation}
    U_X\neq -2X(U_{X\phi}\phi_X+U_{XX}).
\end{equation}
Now, let us suppose the gradient of the scalar field $\xi$ to be timelike, i.e. $\nabla_\mu\xi=a\,t_\mu$, with $g^{\mu\nu}t_\mu t_\nu=-1$ and $a$ some real constant. For this setting we have
\begin{equation}
\Tilde{Q}^{\mu\nu}\nabla_\mu\xi\nabla_\nu\xi=a^2(-1+a^2 K),
\end{equation}
so that $\Tilde{Q}^{\mu\nu}$ is Lorentzian, i.e. $\Tilde{Q}^{\mu\nu}\nabla_\mu\xi\nabla_\nu\xi<0$, if $a^2 K <1$. It is also easy to check that for a unit spacelike vector $s_\mu$, orthogonal to $t_\mu$, the following hold $\Tilde{Q}^{\mu\nu}s_\mu s_\nu=1, \Tilde{Q}^{\mu\nu}s_\mu t_\nu=0$. Conversely, when $\nabla_\mu\xi$ is spacelike, i.e. $\nabla_\mu\xi=a\,s_\mu$, we have
\begin{equation}
\Tilde{Q}^{\mu\nu}\nabla_\mu\xi\nabla_\nu\xi=a^2(1+a^2 K),
\end{equation}
so that by requiring now $\Tilde{Q}^{\mu\nu}\nabla_\mu\xi\nabla_\nu\xi>0$, we end up with the condition $a^2K>-1$. Then, by noticing that for a timelike and a spacelike scalar field gradient we have, respectively, $a^2=\mp X$, we obtain the general condition $KX>-1$, which can be eventually reformulated as
\begin{equation}
   2X(U_{X\phi}\phi_X+U_{XX})+U_X>0.
\end{equation}
Up to a factor in the definition of the kinetic term $X$, this relation retains the same structure of \cite{Babichev:2007dw}, with the additional term $U_{X\phi}\phi_X$. Such an effective metric determines therefore the partition of the local acoustic spacetime into past, future and causally disconnected regions. Indeed, without addressing into details the issue of the characteristic surfaces for the perturbations (see \cite{Babichev:2007dw}), here we just sketch the main idea by considering the acoustic cone defined by the null vectors of the effective metric $\Tilde{Q}_{\mu\nu}^{-1}$. The condition $\Tilde{Q}_{\mu\nu}^{-1}N^\mu N^\nu=0$ can be indeed restated as a relation for the scalar product in the original spacetime metric $g_{\mu\nu}$, that is
\begin{equation}\label{lightcone}
    g_{\mu\nu}N^\mu N^\nu = \frac{K}{1+KX}\leri{N^\mu \nabla_\mu\xi }^2.
\end{equation}
Since an hyperbolic equation of motion requires $1+KX>0$, we see that the sign of $K$ determines if the influence cone is larger ($K>0$) or smaller ($K<0$) than the light cone defined by $g_{\mu\nu}$, opening up to the possibility of having a super-luminal signal velocity for the wave-fronts. In standard $k$-essence models this does not lead to causal inconsistencies \cite{Babichev:2007dw}, due to the fact that acoustic perturbations propagate through an effective medium defined by the non-trivial configuration of the background field, which selects a preferred reference frame breaking Lorentz invariance. We expect the same kind of arguments to hold also in our generalized Palatini formulation, up to possible contributions stemming from the structural equation, and ultimately due to the non-minimal coupling of the matter with the dynamical scalar field $\xi$. This being said, we outline that the particular choice on the potential imposed in \eqref{formau} implies that the metric defining the principal part is conformally related to the spacetime metric, i.e. $Q^{\mu\nu}=\lambda^2 \phi g^{\mu\nu}$. For this configuration, the parameter $K$ results identically vanishing irrespective of the presence of matter, so that, by looking at \eqref{lightcone}, we see that acoustic and light cones coincide. In standard $k$-essence theories this cannot be achieved, since $K=0$ would just imply $U_{XX}=0$, corresponding to a linear dependence of $U$ on $X$. In our case, instead, according the value of the parameter $m$, the potential form \eqref{formau} allows in principle for a wider non-linear dependence of $U$ on $X$, retaining the property $\phi=\phi(\xi,T)$. By setting $m=1$, one can therefore consider the simplest scenario where the $k$-essence field is not trivially coupled to the cosmological fluid, restricting the non-linearities in the derivatives of the dynamical field $\xi$ to quadratic order. As explained in \autoref{app: a}, this property cannot be simply reproduced by considering a Palatini $f(\pal)$ gravity model endowed with a $k$-essence field, in that in this case the equation for $\xi$ is actually insensible of the structural equation. In this sense, the model discussed in \autoref{sec: 4} and displayed by \eqref{formau}, represents the minimal prescription able to encapsulate the effect of a non-minimal coupling between the $k$-essence and the (cosmological) matter field. Interestingly, it can be also viewed as the slow velocity regime of a Dirac-Born-Infeld (DBI) scalar field $\xi$ \cite{Silverstein:2003hf,Alishahiha:2004eh}, exhibiting a coupling with the non-dynamical field $\phi$, i.e.
\begin{equation}
    \mathcal{L}_{DBI} \subseteq \lambda^2\phi \leri{1-\sqrt{1+2  X}} \simeq -\lambda^2\phi X.
\end{equation}
We conclude this section by noticing that the problem of the stability can be equivalently addressed from the perspective of the DHOST theories \cite{BenAchour:2016cay,Langlois:2015cwa}. These represent a class of scalar-tensor theories exhibiting higher order equations of motion, where degeneracy, in the form of additional constraints in the phase space, assures the absence of propagating ghost modes. This analogy can be appreciated by considering our model in vacuum, where, once a formal solution $\phi=\phi(\xi,X)$ is obtained, the initial action \eqref{eq: action f(R)} can be in principle recast as 
\begin{equation}
    S=\jord \leri{\phi R+\frac{3\phi_\xi^2X}{2\phi}-U(\xi,X)+\frac{3\phi_\xi\phi_X}{\phi}\xi^\mu \xi^\nu \xi_{\mu\nu}+\frac{6\phi_X^2}{\phi}\xi^\rho\xi_{\rho\mu}\xi^\mu_{\;\,\sigma}\xi^\sigma}
    \label{eq: eff action dhost}
\end{equation}
with the shortcut notation $\xi_\mu = \nabla_\mu\xi,\;\xi_{\mu\nu}=\nabla_\mu\nabla_\nu\xi$. A direct comparison with \cite{BenAchour:2016cay} allows us then to recognize \eqref{eq: eff action dhost} as an element of the DHOST subclass Ia, up to the identification
\begin{align}
    F(\xi,X)&=\phi\\
    P(\xi,X)&=\frac{3\phi_\xi^2X}{2\phi}-U(\xi,X)\\
    Q_1(\xi,X)&=0\\
    Q_2(\xi,X)&=\frac{3\phi_\xi\phi_X}{\phi}\\
    \alpha_1(\xi,X)&=\alpha_2(\xi,X)=\alpha_3(\xi,X)=\alpha_5(\xi,X)=0\\
    \alpha_4(\xi,X)&=\frac{6\phi_X^2}{\phi}    
\end{align}
where the function $F,P,Q_i,\alpha_i$ are defined in \cite{BenAchour:2016cay}. An analogous result was obtained in \cite{Bombacigno:2021bpk}, and also in the present case the resulting DHOST model is compatible with the phenomenological bound provided by GW170817 \cite{Langlois:2017mxy,Langlois:2017dyl}, demanding that the speed of light coincides with that of gravitational waves.

\section{Dynamical system analysis of the FLRW cosmology}\label{sec: 4}
\noindent In this section we are interested in the dynamics of the system for a cosmological background described by the flat Friedmann–Lemaître–Robertson–Walker (FLRW) metric, whose line element is described in Cartesian coordinates and comoving time by
\begin{equation}
    ds^2=-dt^2+a(t)^2 (dx^2+dy^2+dz^2),
\end{equation}
with $a(t)$ denoting the scale factor. The function carrying the $k$-essence term is chosen as in \eqref{eq: U final}, which as discussed in  \autoref{sec: 3}, automatically satisfies the no-ghost condition. In this setting, the equations for the metric take the form
\begin{align}
&3H^2+3H\frac{\dot{\phi}}{\phi}+\frac{3}{4}\leri{\frac{\dot{\phi}}{\phi}}^2-\frac{1}{2}\lambda^2\dot\xi^2-\frac{W}{2\phi}=\frac{\kappa^2\rho}{\phi}
\label{eq:Friedmann}\\
&-2\dot H-3H^2-2H\frac{\dot{\phi}}{\phi}-\frac{\Ddot{\phi}}{\phi}+\frac{3}{4}\leri{\frac{\dot{\phi}}{\phi}}^2-\frac{1}{2}\lambda^2\dot\xi^2+\frac{W}{2\phi}=\frac{\kappa^2 P}{\phi},
\label{eq:acceleration}
\end{align}
while the evolution of the scalar fields is displayed by
\begin{align}
    &2W-\phi W_\phi=\kappa^2 T
    \label{eq: structural}\\
    &\Ddot{\xi}+\leri{3H+\frac{\dot\phi}{\phi}}\dot \xi+\frac{W_\xi}{2\lambda^2\phi}=0.
    \label{eq: xi frw}
\end{align}
We adopted for the energy momentum tensor the perfect fluid description $ T_{\mu\nu}=\text{diag}(\rho, P, P,P)$ and we introduced the dot notation for derivation with respect to the comoving time. In particular, we used the identity $X=-\dot\xi^2$. 
\\ We note that by introducing the modified Hubble function $Z=H+\frac{1}{2}\frac{\dot{\phi}}{\phi}$, it is possible to eliminate in \eqref{eq: friedman frw z}-\eqref{eq: xi frw z} the second order time derivatives of the field $\phi$. Under such a redefinition, the system takes the form
\begin{align}
    &3Z^2-\frac{1}{2}\lambda^2 \dot{\xi}^2-\frac{W}{2\phi}=\frac{\kappa^2\rho}{\phi}\label{eq: friedman frw z}
\\
    &-2 \dot{Z}-3Z^2+Z \frac{\dot{\phi}}{\phi}-\frac{1}{2}\lambda^2 \dot \xi ^2+\frac{W}{2\phi}=\frac{\kappa^2 w\rho}{\phi} \label{eq: acceleration frw z}
    \\
    &2W-\phi W_\phi=\kappa^2 (3w-1)\rho,\label{eq: structural z}
    \\
    &\Ddot{\xi}+\dot \xi \leri{3Z-\frac{ \dot \phi}{2\phi}}+\frac{W_\xi}{2\lambda^2 \phi}=0,\label{eq: xi frw z}
\end{align}
where we implemented for the pressure the equation of state $P=w\rho$. Then, we define the set of dimensionless dynamical variables
\begin{align}\label{sdv1}
    q&=-\leri{1+\frac{Z'}{Z}}\\
    x&=\frac{\phi'}{\phi}\label{sdv2}\\
    Q&= 2q+x\label{def Q}\\
    y&=\xi'\\
    \Omega &=\frac{\kappa^2\rho}{3\phi Z^2}\label{sdv3}\\
    u&=\frac{W}{6\phi Z^2}\label{def u}\\
    u_\phi&=\frac{W_\phi}{3 Z^2}\label{def uphi}\\
    u_\xi&=\frac{W_\xi}{2\lambda^2\phi Z^2}\label{def uxi}
\end{align}
with a prime denoting differentiation w.r.t. the dimensionless time $\tau\equiv \ln (\sqrt{\phi}a)$. It is hence possible to rewrite the system \eqref{eq: friedman frw z}-\eqref{eq: xi frw z} as
\begin{align}
    &1-\frac{\lambda^2}{6}y^2-u=\Omega\label{eq: dyn syst 1}\\
    &\frac{Q}{3}-\frac{1}{3}-\frac{\lambda^2}{6}y^2+u=w\Omega\label{eq: dyn syst 2}\\
    &4u-u_\phi=(3w-1)\Omega\label{eq: dyn syst 3}\\
    &y'=y\leri{\frac{Q}{2}-2}-u_\xi\label{eq: dyn syst 4}.
\end{align}
We see that \eqref{eq: dyn syst 1}-\eqref{eq: dyn syst 3} actually constitute a set of algebraic constraints, allowing us to solve for $\leri{u,Q,u_\phi}$ in terms of $\leri{y,\Omega}$, i.e.
\begin{align}
    &u=1-\frac{\lambda^2 y^2}{6}-\Omega
    \label{eq: sol u}\\
    &Q=-2+\lambda^2y^2+3(1+w)\Omega
    \label{eq: sol Q}\\
    &u_\phi=4-\frac{2\lambda^2y^2}{3}-3(1+w)\Omega
    \label{eq: sol u phi}.
\end{align}
The only truly dynamical equation is then displayed by \eqref{eq: dyn syst 4}, which besides $\leri{y,\Omega}$ also contains the variable $u_\xi$. This implies that in order the dynamical system be well posed, we need two additional dynamical relations for $\Omega$ and $u_\xi$. For the same reason, we expect that the variable $x$ could be expressed algebraically in terms of the set $\leri{y,\Omega,u_\xi}$. That is consistent with the number of the original degrees of freedom, i.e. the scale factor $a(t)$, the scalar field $\xi(t)$ and the matter content $\rho(t)$. The equation for $\Omega$ can be determined starting from its definition, i.e.
\begin{equation}
    \begin{split}
        \dot\Omega&=\frac{\kappa^2\dot\rho}{3\phi Z^2}-\frac{\kappa^2\rho\dot\phi}{3\phi^2 Z^2}-\frac{2\kappa^2\rho \dot Z}{3\phi Z^3}=\\
        &=\frac{\kappa^2\dot\rho}{3\phi Z^2}-\frac{\kappa^2\rho}{3\phi Z}\frac{\phi'}{\phi}-\frac{2\kappa^2\rho}{3\phi Z}\frac{Z'}{Z}=\\
        &=\frac{\kappa^2\dot\rho}{3\phi Z^2}-Z\Omega\leri{x-2q-2},
    \end{split}
\end{equation}
which can be solved for $\dot\rho$ leading to
\begin{equation}
    \frac{\kappa^2\dot\rho}{3\phi Z^2}=Z\Omega'+Z\Omega\leri{x-2q-2}.
\end{equation}
The continuity equation $\dot\rho+3H(1+w)\rho=0$ can be rewritten as
\begin{align}
    \frac{\kappa^2\dot\rho}{3\phi Z^2}=-3Z(1+w)\leri{1-\frac{x}{2}}\Omega,
\end{align}
so that by comparison we eventually obtain
\begin{equation}
    \Omega'=\Omega\leri{Q-\frac{x}{2}-1-3w\leri{1-\frac{x}{2}}}.
    \label{eq: omega}
\end{equation}
Now, in order to derive an expression for $x$, we start by observing that \eqref{def u}-\eqref{def uxi} allow to express the set $\leri{\phi,\xi,Z^2}$ in terms of the variables $\leri{u,u_\phi,u_\xi}$. In particular, in the following we will assume that the potential is factorizable as
\begin{equation}
    W(\phi,\xi)=W_0 g(\phi) f(\xi),
\end{equation}
where $W_0$ is a constant of dimension of $L^{-2}$. With this choice we obtain
\begin{align}
    &\frac{u_\phi}{2u}=\frac{W_\phi \phi}{W}=\frac{\phi g_\phi(\phi)}{g(\phi)}\label{eq: inv g}\\
    &\frac{\lambda^2}{3}\frac{u_\xi}{u}=\frac{W_\xi}{W}=\frac{f_\xi(\xi)}{f(\xi)}\label{eq: inv f},
\end{align}
where we note that the right hand members are function only of $\phi$ and $\xi$, respectively. This pair of equations allows us to solve for such a set of variables, so that taking into account \eqref{def u}, we eventually obtain the formal expression for $Z^2$, i.e.
\begin{equation}
    Z^2=\frac{W(\phi(u,u_\phi),\xi(u,u_\xi))}{6u\phi(u,u_\phi)}.\label{eq: inv Z}
\end{equation}
We remark that the choice of a factorizable potential is actually not mandatory to obtain a well-defined inversion scheme. Other forms are indeed feasible, involving for instance mixed polynomials in the fields or more exotic functional expressions, provided that \eqref{eq: inv g}-\eqref{eq: inv f} can be explicitly solved for $\phi$ and $\xi$. When the factorizable potential case is addressed, however,  we have to require
\begin{equation}
    g(\phi)\neq g_0 \phi^k,\qquad f(\xi)\neq f_0 e^{k\xi},
\end{equation}
with $g_0,f_0,k$ some constants, in order \eqref{eq: inv g}-\eqref{eq: inv f} can be solved explicitly. Be this the case, indeed, \eqref{eq: inv g}-\eqref{eq: inv f} reduce to identities and the system is underdetermined. Now, if we take the definition of $x$ and we consider $\phi$ as a function of $\leri{u,u_\phi}$, i.e. as a function of $\leri{y,\Omega}$, we end up with
\begin{equation}
    x=\frac{\phi'(y,\Omega)}{\phi(y,\Omega)}=\frac{\phi_y y'+\phi_\Omega \Omega'}{\phi(y,\Omega)}.\label{eq: def sol x}
\end{equation}
Since as it can be appreciated from \eqref{eq: omega}, $\Omega'$ also depends on $x$, once we substitute \eqref{eq: omega} in \eqref{eq: def sol x} we need to solve algebraically for $x$. Calculations then show that
    \begin{equation}
    x=\frac{\phi_y(y,\Omega)\lerisq{\frac{y}{2}\leri{-6+\lambda^2 y^2+3(1+w)\Omega}-u_\xi}+\phi_\Omega(y,\Omega)\,\Omega\lerisq{\lambda^2 y^2+3(1+w)(\Omega-1)}}{\phi(y,\Omega)+\frac{1-3w}{2}\,\phi_\Omega(y,\Omega)\,\Omega }.
    \label{eq: solution x}
\end{equation}
Eventually, we proceed by evaluating the equation for the $u_\xi$ variable, obtaining
\begin{equation}
    u'_\xi=\frac{1}{2\lambda^2}\leri{\Gamma_{\phi\xi}x+\Gamma_{\xi\xi}y}+2u_\xi \leri{Q-x+1},
    \label{eq: dyn uxi}
\end{equation}
where we defined in analogy with the existing literature \cite{Bahamonde:2017ize} the quantities
\begin{align}
    &\Gamma_{\phi\xi}\equiv \frac{ W_{\phi\xi}}{Z^2}=\lambda^2\frac{u_\phi u_\xi}{u}\\
    &\Gamma_{\xi\xi}\equiv \frac{ W_{\xi\xi}}{\phi Z^2}=\frac{6u}{f(\xi(u,u_\xi))}\frac{d^2 f(\xi(u,u_\xi))}{d \xi^2}.\label{gammaxixi}
\end{align}
We note that once \eqref{eq: inv g}-\eqref{eq: inv f} are solved for $\phi$ and $\xi$, respectively, $\Gamma_{\phi\xi}$ and $\Gamma_{\xi\xi}$ are completely determined in terms of $(u,u_\phi,u_\xi)$, namely in terms of the independent variables $(y,\Omega,u_\xi)$, whose dynamical equations we report here all at once for the sake of clarity
\begin{equation}
   \begin{cases}
    &y'=y\leri{\frac{Q}{2}-2}-u_\xi\\
    &\Omega'=\Omega\leri{Q-\frac{x}{2}-1-3w\leri{1-\frac{x}{2}}}\\
    &u'_\xi=\frac{1}{2\lambda^2}\leri{\Gamma_{\phi\xi}x+\Gamma_{\xi\xi}y}+2u_\xi \leri{Q-x+1}.
    \label{eq: dyn syst}
\end{cases} 
\end{equation}
We observe that \eqref{eq:Friedmann} and \eqref{eq:acceleration} can be also recast in the suggestive form
\begin{equation}
    H^2=\frac{\kappa^2 \rho_{\text{eff}}}{3\phi},\qquad-2\dot H-3H^2=\frac{\kappa^2 P_{\text{eff}}}{\phi}
\end{equation}
where the effective energy density and pressure have been introduced, i.e.
\begin{align}
    &\rho_{\text{eff}}=\rho+\frac{1}{\kappa^2}\leri{-3Z \dot\phi+\frac{3\dot\phi^2}{\phi}+\frac{1}{2}\lambda^2 \phi \dot\xi^2+\frac{W}{2}}\\
    &P_{\text{eff}}=P+\frac{1}{\kappa^2}\leri{2Z \dot\phi-\frac{7\dot\phi^2}{4\phi}+\ddot \phi+\frac{1}{2}\lambda^2 \phi \dot\xi^2-\frac{W}{2}}.
\end{align}
That allows to define the effective barotropic index $w_{\text{eff}}=\frac{P_{\text{eff}}}{\rho_{\text{eff}}}$, whose expression in terms of the dynamical variables is displayed by
\begin{equation}
    w_{\text{eff}}=\frac{w \Omega+\frac{1}{3}\leri{q-1}x-\frac{1}{4}x^2+\frac{1}{3}x'+\frac{1}{6}\lambda^2 y^2-u}{\Omega-x+\frac{1}{4}x^2+\frac{1}{6}\lambda^2y^2+u},
    \label{eq: eff w}
\end{equation}
where the derivative of $x$ has to be understood as
\begin{equation}
    x'=x_y y'+x_\Omega \Omega'+x_{u_\xi} u_\xi'.
\end{equation}
We conclude this section by discussing some explicit forms for the function $g(\phi)$ and $f(\xi)$, defining the potential $W(\phi,\xi)$, which allow for the inversion of \eqref{eq: inv f}-\eqref{eq: inv g}. In respect to $g(\phi)$ we choose two different cases of interest, given by the exponential and power-law behaviors, i.e.
\begin{align*}
    g(\phi)&=e^{a\phi^k},\quad g(\phi)=\leri{\phi-\phi_0}^p,
\end{align*}
resulting, respectively, in the following expressions for the field $\phi$:
\begin{align}
    \phi&=
    \leri{\frac{u_\phi}{2aku}}^{\frac{1}{k}},
    %=\lerisq{\frac{1}{ak}\leri{2+\frac{3(1-3w)\Omega}{6-\lambda^2 y^2-6\Omega}}}^{\frac{1}{k}}\\
    \quad\phi
    =\frac{\phi_0 u_\phi}{u_\phi-2pu},
    %=\phi_0\leri{\frac{2(6-\lambda^2 y^2-6\Omega)+3(1-3w)\Omega}{(2-p)(6-\lambda^2 y^2-6\Omega)+3(1-3w)\Omega}}.
\end{align}
Analogously, for $f(\xi)$ we restrict our attention to
\begin{align*}
    &f(\xi)=e^{a\xi+b\xi^k},\\
    &f(\xi)=(\xi-\xi_0)^n,
\end{align*}
which once plugged back in \eqref{eq: inv f} lead to
\begin{align}
\xi=&\lerisq{\frac{1}{kb}\leri{\frac{\lambda^2}{3}\frac{u_\xi}{u}-a}}^{\frac{1}{k-1}},\quad
%&\lerisq{\frac{1}{kb}\leri{\frac{2\lambda^2 u_\xi}{6-\lambda^2 y^2-6\Omega}-a}}^{\frac{1}{k-1}},\\
\xi=\;\xi_0+\frac{3nu}{\lambda^2 u_\xi}.
%\xi_0+\frac{n}{2\lambda^2}\frac{6-\lambda^2 y^2-6\Omega}{u_\xi}.
\end{align}
We remind the reader that in the expression of $\phi,\xi$ the variables $u,u_\phi$ must be understood in terms of $y,\Omega$, as given by \eqref{eq: sol u} and \eqref{eq: sol u phi}.
In the next sections, thus, we will specialize the analysis to two specific scenarios, where the global potential $W(\phi,\xi)$ is a product of either exponential or power-law profiles, i.e.
 \begin{align}
      W&=W_0 \, e^{a\phi}e^{b\xi^2}\\
      W&=W_0 \leri{\phi-\phi_0}^p \leri{\xi-\xi_0}^k.
      \label{eq: potential forms}
 \end{align}
In particular, in  \autoref{sec: 5} we will consider a two-component $T_{\mu\nu}$ description taking into account standard matter and radiation cosmological fluids, and we will look at how the value of the parameters defining the potential affects the nature of the solutions. Then, in  \autoref{sec: 6}, we will assume instead a single fluid contribution, keeping the barotropic index undetermined.\\

\section{Two-fluids configuration}\label{sec: 5}
Here we consider the case where the energy momentum tensor carries two different non interacting fluids, represented respectively by matter ($w=0$) and radiation ($w=1/3$). For this setting, every fluid has its proper dynamical equation \eqref{eq: omega}, i.e.
\begin{equation}
    \Omega_M'=\Omega_M\leri{Q-\frac{x}{2}-1},\qquad\Omega_R'=\Omega_R\leri{Q-2},
    \label{eq: omega R M}
\end{equation}
and the effective barotropic index takes the form
\begin{equation}
     w_{\text{eff}}=\frac{\frac{1}{3}\Omega_R+\frac{1}{3}\leri{q-1}x-\frac{1}{4}x^2+\frac{1}{3}x'+\frac{1}{6}\lambda^2 y^2-u}{\Omega_M+\Omega_R-x+\frac{1}{4}x^2+\frac{1}{6}\lambda^2y^2+u},
\end{equation}
with
   \begin{multline}
       x=\frac{1}{\phi+\frac{1}{2}\Omega_M\phi_M}\bigg [ \phi_y\lerisq{\frac{y}{2}\leri{-6+\lambda^2 y^2+3\Omega_M+4\Omega_R}-u_\xi}+\\
+\Omega_M\phi_M\lerisq{\lambda^2 y^2+3\Omega_M+4\Omega_R-3}+\Omega_R\phi_R\lerisq{\lambda^2 y^2+3\Omega_M+4\Omega_R-4} \bigg],
   \end{multline}
where the subscript $M,R$ on $\phi$ denotes derivation with respect to the corresponding component of the fluid, and the independent dynamical variables are displayed by the four quantities $(y,\Omega_M,\Omega_R,u_\xi)$.
\subsection{Exponential potential}\label{2fluexp}
In this section we study the evolution of the dynamical system under the influence of the potential
\begin{equation}
    W=W_0 \, e^{a\phi}e^{b\xi^2}.
\end{equation}
where $(a,b)$ are two real numbers. Under this choice the scalar fields are obtained as 
\begin{equation}
    \phi=\frac{u_\phi}{2au} \qquad \xi=\frac{\lambda^2 u_\xi}{6bu}, \label{scfieldsdoppioexp}
\end{equation}
while the quantity $\Gamma_{\xi\xi}$ entering the dynamical equation for the variable $u_\xi$ reads
\begin{equation}
   \Gamma_{\xi\xi}=12bu+\frac{2\lambda^4 u_\xi^2}{3u}.
\end{equation}
It is clear that the potential, given its exponential dependence from the scalar fields, is compelled to be positive, therefore we have to restrict the possible values of the 4D vector of dynamical variables $(y,\Omega_M,\Omega_R,u_\xi)$ in order to have $u>0$. The same applies to the field $\phi$, which we require to satisfy $\phi>0$ for preserving the attractive character of gravity. Last, the matter and radiation densities $\Omega_M$ and $\Omega_R$ are constrained to be non-negative and the other dynamical variables $y$ and $u_\xi$ must always remain real. The conditions just enumerated are necessary and sufficient for the existence of a fixed point of the considered dynamical system, which explicitly reads   
\begin{equation}
   \begin{cases}
    &y'=y\leri{\frac{Q}{2}-2}-u_\xi\\
    &\Omega_M'=\Omega_M\leri{Q-\frac{x}{2}-1} \\
    &\Omega_R'=\Omega_R\leri{Q-2}\\
    &u'_\xi=\frac{1}{2\lambda^2}\leri{\Gamma_{\phi\xi}x+\Gamma_{\xi\xi}y}+2u_\xi \leri{Q-x+1},
    \label{eq: dyn syst two fluids}
\end{cases} 
\end{equation}
and will be enforced as a selection criterion to discard non-physical regions of the phase space. This being said, we find that the dynamical system admits up to $7$ fixed points:
\begin{align}
    & P_1=\leri{0,0,0,0}\\
    &P_{2,3}= \leri{\pm\frac{\sqrt{2\leri{b+\lambda^2}}}{\lambda^2},0,0,\pm\frac{\leri{b-2\lambda^2}\sqrt{2\leri{b+\lambda^2}}}{\lambda^4}} \\
    &P_{4,5}= \leri{\pm\frac{2}{\sqrt{b}},0,1-\frac{\lambda^2}{b} ,\mp\frac{2}{\sqrt{b}}}\\
&P_{6,7}=\leri{\pm\sqrt{\frac{3}{2b}},1-\frac{\lambda^2}{2b},0,\mp\frac{3\sqrt{3}}{2\sqrt{2b}}}.
\end{align}
In order to study the existence of such points it is useful to introduce the parameterization $b=s \lambda^2$ with $s$ a real number. Clearly the value $s=0$ must be excluded from the analysis, for it implies a divergent scalar field $\xi$, as it can be appreciated by the inspection of \eqref{scfieldsdoppioexp}. Beside this, it is found that the origin $P_1$ is a fixed point regardless of the choice of $s$, while the other vacuum solutions corresponding to the points $P_{2,3}$ exist for $-1 \leq s <2 $. In the limit $s\to -1$ both fixed points $P_{2,3}$ collapse in the origin, becoming coincident with $P_1$. For what concerns the points $P_{4,5}$ and $P_{6,7}$ we calculate interval of existence given by $s \geq 1$ and $s \geq \frac{1}{2}$, respectively. 
It has to be remarked that the value of the scalar field $\phi$ in the points $P_i$ with $i=1,\dots,5$ is $\frac{2}{a}$, whereas in $P_{6,7}$ results $\phi=\frac{1+2s}{a}$, therefore the supplemental condition $a>0$ has to be imposed to guarantee the existence of the entire set of fixed points. 
Next, we proceed to study the stability, i.e. the character of the local dynamics in the neighborhood of the fixed points. For this purpose, we evaluate the Jacobian matrix encoding the first derivatives of the right-handed side of the system \eqref{eq: dyn syst two fluids} with respect to the dynamical variables and we calculate the eigenvalues corresponding to each fixed point, whose explicit expression is reported in Appendix \ref{app:b}. It is found that:
\begin{itemize}
    \item For $s>-1$ the point $P_1$ results to be an attractor, i.e. all the eigenvalues have negative real part, whereas for $s<-1$ we outline the presence of at least two eigenvalues with real part having opposite sign. Hence, for $s<-1$ the point $P_1$ is a saddle. When $s=-1$, instead, the analysis of the Jacobian matrix eigenvalues fails to give a definite outcome. Indeed, in this case it is found that three eigenvalues have negative real part while the other is exactly null. We proceed to evaluate the stability of the point $P_1$ for $s=-1$ by making use of center manifold theory. As a first step we write the system \eqref{eq: dyn syst two fluids} in a set of coordinates with respect to which the first order dynamics is diagonal, namely
    \begin{equation}
    \begin{split}
             &q_1=\Omega_M \qquad \qquad
             \quad \; \,
             q_2=\Omega_R\\
             &q_3=\frac{2}{5}\leri{3y+u_\xi} \qquad q_4=\frac{3}{5}\leri{-2y+u_\xi}.
    \end{split}
    \end{equation}
    For this set of coordinates the dynamical system is written as 
    \begin{equation}
   \begin{cases}
    &q_1'=-3q_1+f_1\leri{q_1,q_2,q_3,q_4}\\
    &q_2'=-4q_2+f_2\leri{q_1,q_2,q_3,q_4} \\
    &q_3'=-5q_3+f_3\leri{q_1,q_2,q_3,q_4}\\
    &q_4'=g\leri{q_1,q_2,q_3,q_4}
\end{cases} 
\end{equation}
where the functions $f_i$ and $g$ encode the non-linearities of the dynamical equations. Next we define three functions of the center manifold coordinate $q_4$, for which we assume a Taylor expansion of the form
\begin{align}
    &h_1\leri{q_4}=\alpha_2 q_4^2+\alpha_3 q_4^3+\alpha_4 q_4^4 \\
     &h_2\leri{q_4}=\beta_2q_4^2+\beta_3 q_4^3+\beta_4 q_4^4\\
      &h_3\leri{q_4}=\gamma_2 q_4^2+\gamma_3 q_4^3+\gamma_4 q_4^4.
\end{align}
The coefficients characterizing the functions $h_i$ are found from
\begin{equation}\label{detcenter}
    h_i' \leri{q_4} g\leri{h_i,q_4}-A_{ij}h_j \leri{q_4}-f_i\leri{h_i,q_4}=0,
\end{equation}
where $A_{ij}$ are the components of the matrix $A=\text{diag}\leri{-3,-4,-5}$. An algebraic system for the coefficients of the Taylor expansion is obtained, giving
\begin{equation}
\begin{split}
    &\alpha_2=\alpha_3=\alpha_4=0\\
    &\beta_2=\beta_3=\beta_4=0\\
    &\gamma_2=\gamma_4=0\\
    &\gamma_3=\frac{\lambda^2}{675}.
\end{split}
\end{equation}
The local dynamics of the coordinate $q_4$ restricted to the center manifold is then described by the dynamical equation
\begin{equation}
    q_4'=\frac{\lambda^2 q_4^3}{15}+\mathcal{O}\leri{q_4^5},
\end{equation}
predicting instability in the direction relative to the coordinate $q_4$. Therefore, for $s=-1$ the point $P_1$ results to be a saddle. In  \autoref{pporigineattr} we draw phase portraits of the neighborhood of the point $P_1$ when the latter is an attractor for the system, assuming $s=\frac{3}{2}$.

\begin{figure*}[htbp!]
\captionsetup{justification=justified, singlelinecheck=false}
\centering
\begin{subfigure}{0.45\textwidth}
\centering
\includegraphics[width=\linewidth]{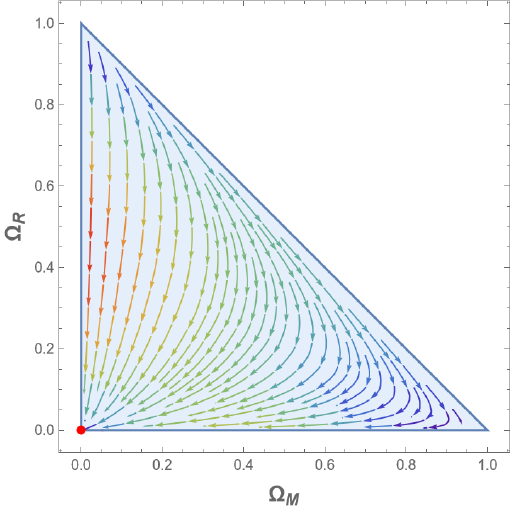}
\end{subfigure}
\hfill
\begin{subfigure}{0.45\textwidth}
\centering
\includegraphics[width=\linewidth]{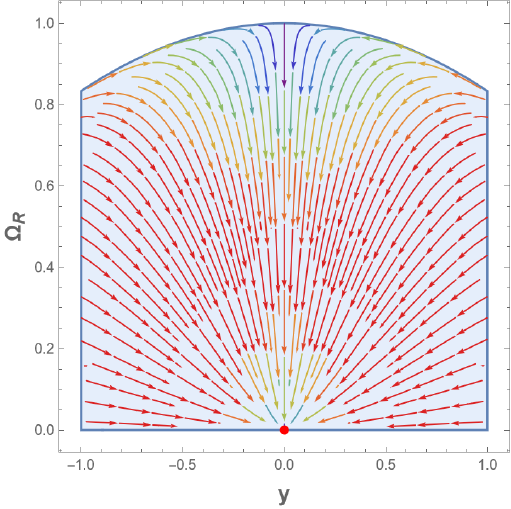}
\end{subfigure}

\vspace{0.4cm}

\begin{subfigure}{0.45\textwidth}
\centering
\includegraphics[width=\linewidth]{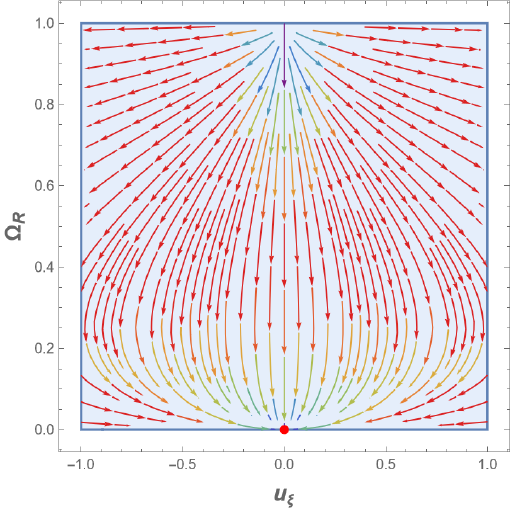}
\end{subfigure}
\hfill
\begin{subfigure}{0.45\textwidth}
\centering
\includegraphics[width=\linewidth]{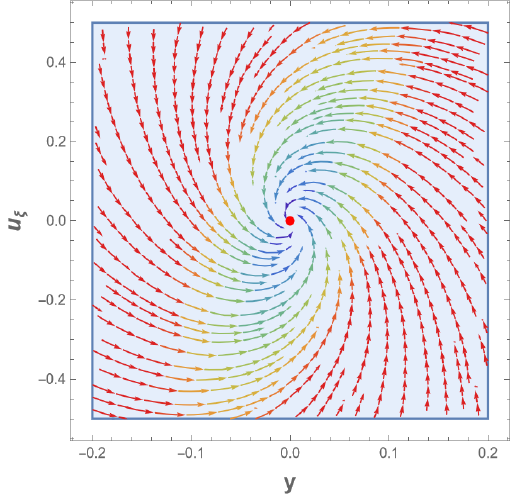}
\end{subfigure}
\caption{Phase portraits of several cross-sections in the neighborhood of the point $P_1$ (red dot) when the latter is an attractor ($a=1$, $s=\frac{3}{2}$, $\lambda=1$, $W_0=1$). The colors of the arrows indicate different magnitudes of the derivatives in each point, growing from blue to red. In light blue the permitted region of the phase space as it is defined by the constraint $u=1-\frac{\lambda^2 y^2}{6}-\Omega_M-\Omega_R>0$.}
\label{pporigineattr}
\end{figure*}

In  \autoref{pporiginesaddle} we report instead the cross-section of the phase space corresponding to the $\leri{y,u_\xi}$ plane when the point $P_1$ results to be a saddle for the dynamical system, setting $s=-2$. We omit to display the other cross-sections which remain qualitatively similar to the previous case.
\begin{figure}[htbp]
\captionsetup{justification=justified, singlelinecheck=false}
    \centering
\includegraphics[width=7cm]{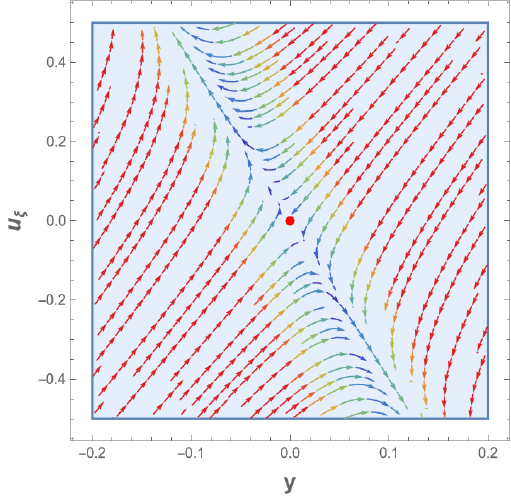}
    \caption{Phase portrait of the point $P_1$ for $a=1$, $s=-2$, $\lambda=1$, $W_0=1$. We display only the cross-section relative to the $\leri{y,u_\xi}$ plane, where the saddle nature of such point is manifest.}
    \label{pporiginesaddle}
\end{figure}
\item The points $P_{2,3}$ are always saddles in their range of existence, namely $-1 \leq s < 2$ and $s\neq 0$.  In  \autoref{ppp2} we display cross-sections of the phase space depicting phase portraits of the neighborhood of the point $P_2$, setting $s=\frac{3}{2}$. As it can be observed trajectories are either repelled or attracted to the fixed point depending on the specific direction considered. 

\begin{figure}[htbp]
\captionsetup{justification=justified, singlelinecheck=false}
\centering
\begin{subfigure}{0.45\textwidth}
\centering
\includegraphics[width=\linewidth]{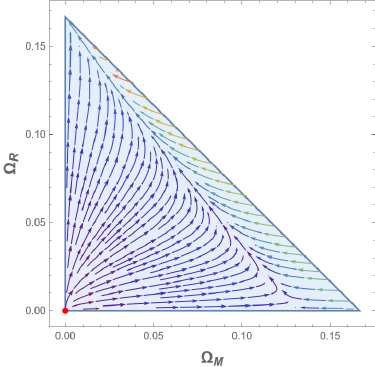}
\end{subfigure}
\hfill
\begin{subfigure}{0.45\textwidth}
\centering
\includegraphics[width=\linewidth]{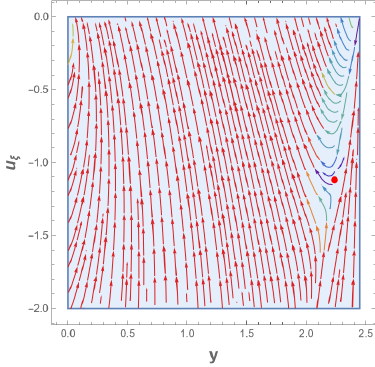}
\end{subfigure}
\caption{Phase portraits of the neighborhood of the point $P_2$ (red dot) for $a=1$, $s=\frac{3}{2}$, $\lambda=1$, $W_0=1$. In light blue the permitted region of the phase space from the constraint $u=1-\frac{\lambda^2 y^2}{6}-\Omega_M-\Omega_R>0$. The colors of the arrows indicate different magnitudes of the derivatives in each point, growing from blue to red.}
\label{ppp2}
\end{figure}
The dependence of the real part of the eigenvalues $a_i$ with respect to the parameter $s$ is depicted in  \autoref{autovaloridoppioexp}.
As it can be easily seen, for every value of $s$ there is always at least a couple of eigenvalues having real parts of opposite sign.

It must be remarked that heteroclinic trajectories connecting the points $P_{2,3}$ to $P_1$ can be numerically found. Indeed, starting from an initial datum conveniently chosen it can be seen that the trajectory calculated with a forward-in-time integration ends in $P_1$, whereas a backward integration gives either $P_2$ or $P_3$ (depending on the choice of the initial datum) as source of the motion. We display this finding in  \autoref{eteroclinella}.
\begin{figure}[ht]
\captionsetup{justification=justified, singlelinecheck=false}
    \centering
\includegraphics[width=7cm]{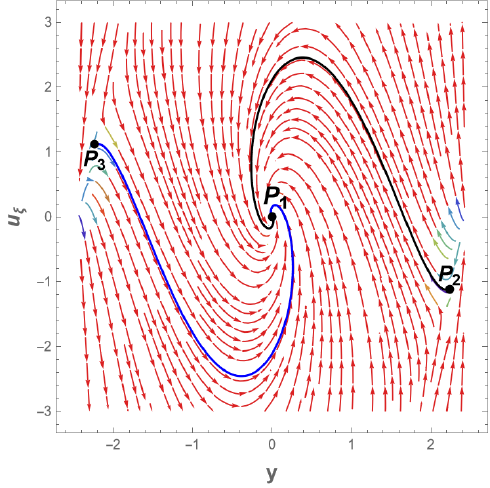}
    \caption{Heteroclinic trajectories connecting the points $P_{2,3}$ with $P_1$. In black (blue) the trajectory starting from $P_2$ ($P_3$). The selected values of the parameters are $a=1$, $s=\frac{3}{2}$, $\lambda=1$, $W_0=1$.}
    \label{eteroclinella}
\end{figure}
We outline that for $s=-1+\varepsilon$, with $\varepsilon$ a small positive number, the physics associated to the points $P_{4,5}$ corresponds to a quasi-de Sitter expansion (see below), whereas in this regime $P_1$ is an attractor for the system and it describes a pure de Sitter phase. Therefore the heteroclinic orbits just described display the existence of a possible transition between two phases of accelerated expansion, potentially reproducing both inflation and late-time de Sitter expansion.
\item The points $P_{4,5}$, which are present for $s\geq 1$, are always saddles, as it can be noticed from the inspection of the plot reported in  \autoref{autovaloridoppioexp}. 

\begin{figure}[htbp!]
\captionsetup{justification=justified, singlelinecheck=false}
\centering
\begin{subfigure}{0.45\textwidth}
\centering
\includegraphics[width=\linewidth]{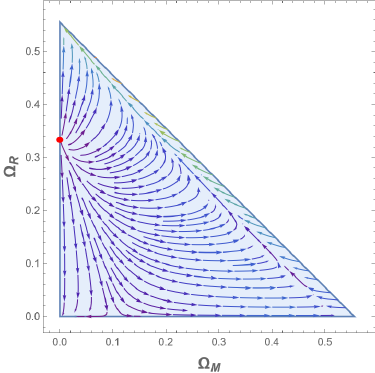}
\end{subfigure}
\hfill
\begin{subfigure}{0.45\textwidth}
\centering
\includegraphics[width=\linewidth]{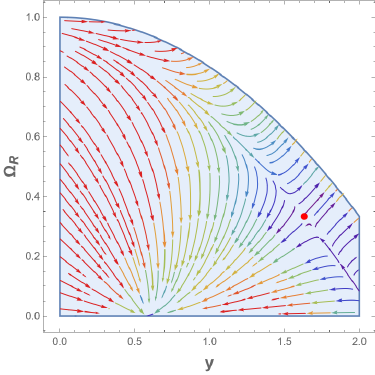}
\end{subfigure}
\caption{Phase portraits of the neighborhood of the point $P_4$ (red dot) for $a=1$, $s=\frac{3}{2}$, $\lambda=1$, $W_0=1$. In light blue the permitted region of the phase space from the constraint $u=1-\frac{\lambda^2 y^2}{6}-\Omega_M-\Omega_R>0$. The colors of the arrows indicate different magnitudes of the derivatives in each point, growing from blue to red.}
\label{ppp4}
\end{figure}
Also in this case we display, in  \autoref{ppp4}, cross-sections of the phase space in the proximity of the point $P_4$, showing the local dynamics typical of a saddle.

\item The points $P_{6,7}$ are saddles in their range of existence, i.e. $s \geq \frac{1}{2}$. The dependence, with respect to $s$, of the Jacobian matrix eigenvalues real part calculated in these points overlaps with great accuracy the corresponding curves relative to the points $P_{4,5}$ plotted in  \autoref{autovaloridoppioexp}. In this case we omit to display phase portraits near the points, given the substantial similarity with the ones reported in  \autoref{ppp4}.
\end{itemize}
After having presented the stability analysis, we are now interested in describing the behavior of the scale factor for each fixed point of the system. The general strategy is to calculate the parameters $q$ and $x$ from the dynamical variables value in the fixed point of interest. This permits to calculate $Z(t)$ and $\phi(t)$ through integrations, hence reconstructing the Hubble parameter $H(t)$. It is found that $x=0$ in all fixed points, then $\phi$ is constant in time and $H(t)=Z(t)$. The parameter $q$, instead, acquires different values depending on the specific fixed point considered. Specifically:
\begin{itemize}
\item In $P_1$ we calculate $q=-1$, hence
\begin{equation}
   \frac{Z'}{Z}=\frac{\dot Z}{Z^2}=0 \quad \Rightarrow \quad Z(t)=c_1 \qquad c_1 \in \mathbb{R}.
\end{equation}
    Given that $H(t)=Z(t)$, this point corresponds to either a de Sitter or an anti de Sitter phase, depending on the sign of the integration constant $c_1$. As a confirmation of this, the effective barotropic index calculated in $P_1$ yields $w_{\text{eff}}=-1$. 
    \item In the points $P_{2,3}$ we calculate $q=s$ (we recall that these points exist for $-1 \leq s <2$ and $s \neq 0$), implying $\frac{\dot Z}{Z^2}=-\leri{s+1}$, from which we evaluate
    \begin{equation}
     Z(t)=\frac{1}{\leri{s+1}t+c_1} \qquad c_1 \in \mathbb{R}.
 \end{equation}
 Therefore, for $s=-1$ we describe again a de Sitter (or anti de Sitter) phase, whereas for $s \neq -1$ we obtain for the scale factor 
 \begin{equation}
    a(t)=C_2 \leri{\leri{s+1}t+c_1}^{\frac{1}{s+1}} \qquad C_2>0.
\end{equation}
Given that $\frac{1}{s+1}$ is always positive in the interval of existence of $P_{2,3}$, we observe that these points correspond to a spacetime expansion, which is either accelerated or decelerated depending on the specific value of the parameter $s$. Concretely, by evaluating the effective barotropic index, which results $w_{\text{eff}}=\frac{2s-1}{3}$, we can see that the points $P_{2,3}$ correspond to accelerated expansions for $-1 \leq s <0$, while for $0<s<2$ we have decelerated expansions with $-\frac{1}{3}<w_{\text{eff}}<1$.
\item For what concerns the points $P_{4,5}$ we calculate $q=1$, which gives for $Z$ the differential equation $\frac{\dot Z}{Z^2}=-2$, whose solution reads
 \begin{equation}
     Z(t)=\frac{1}{2t+c_1} \qquad c_1 \in \mathbb{R}.
 \end{equation}
The scale factor is then readily obtained
\begin{equation}
    a(t)=C_2 \leri{2t+c_1}^{\frac{1}{2}} \qquad C_2>0,
\end{equation}
together with the effective barotropic index, which in the considered fixed points results $w_{\text{eff}}=\frac{1}{3}$. Hence, it is straightforward to recognize that the physics associated to these phase space points is a radiation-dominated decelerated expansion. 

\item Lastly, we consider the points $P_{6,7}$, in which $q=\frac{1}{2}$. Integrating in time the differential equation for $Z$ we obtain
 
 \begin{equation}
     Z(t)=\frac{1}{\frac{3}{2}t+c_1} \qquad c_1 \in \mathbb{R}.
 \end{equation}
Then we calculate the scale factor through another integration, resulting in
\begin{equation}
    a(t)=C_2 \leri{\frac{3}{2}t+c_1}^{\frac{2}{3}} \qquad C_2>0.
\end{equation}
In this case we deal with a matter-dominated decelerated expansion, as also confirmed by the value acquired by the effective barotropic index, namely $w_{\text{eff}}=0$.
\end{itemize}

\subsection{Power-law potential}\label{powerlaw}
In this section we assume a potential of the form
\begin{equation}
    W=W_0 \leri{\phi-\phi_0}^p \leri{\xi-\xi_0}^k 
\end{equation}
in which $p$ and $k$ are integers numbers, while $\phi_0$ and $\xi_0$ are reals with $\phi_0>0$. In this case we evaluate the scalar fields from \eqref{eq: inv g} and \eqref{eq: inv f} as
\begin{equation}
    \phi=\frac{\phi_0 u_\phi}{u_\phi-2p u} \qquad \xi=\xi_0+\frac{3ku}{\lambda^2 u_\xi},
\end{equation}
whereas the quantity $\Gamma_{\xi\xi}$, calculated from \eqref{gammaxixi} reads
\begin{equation}
    \Gamma_{\xi\xi}=\frac{2\leri{k-1}\lambda^4 u_\xi^2}{3ku}.
\end{equation}
In order to guarantee the finiteness of the quantities above, we enumerate the following selection criteria for the existence of fixed points:
\begin{equation}
    \begin{cases}
        &u\neq 0 \\
        &\frac{u_\phi}{u_\phi-2pu}>0\\
        &u_\xi \neq 0
    \end{cases}
\end{equation}
together with the request of having $\Omega_M$ and $\Omega_R$ non-negative and $y$ real. By enforcing these constraints, it turns out that the dynamical system admits a couple of fixed points, namely
\begin{equation}
    P_{1,2}=\leri{\pm\sqrt{\frac{2k}{\lambda^2 \leri{k+1}}},0,0,\mp\sqrt{\frac{2k}{\lambda^2}}\frac{2k+3}{\leri{k+1}^{\frac{3}{2}}} } ,
\end{equation}
for which the existence condition is summarized by $p\leq 1$, $k\neq -1$. In addition to this we observe that the effective barotropic index $w_{\text{eff}}$ diverges for $k=0$, therefore the additional constraint $k \neq 0$ must be imposed. We proceed with the standard stability analysis through the evaluation of the Jacobian matrix eigenvalues calculated in the fixed points, whose explicit expressions in terms of the parameter $k$ read
\begin{align}
    &a_1= -2\frac{k+2}{k+1}\\
    &a_2=-\frac{k+3}{k+1} \\
    &a_3=-4 \\
    &a_4= 2+\frac{1}{k+1}  .
\end{align}
We observe that, for any allowed value of $k$, there are at least two eigenvalues with real part having opposite sign, hence the points $P_{1,2}$ are saddles.

\begin{figure*}[hbtp!]
\captionsetup{justification=justified, singlelinecheck=false}
\centering
\begin{subfigure}{0.45\textwidth}
\centering
\includegraphics[width=\linewidth]{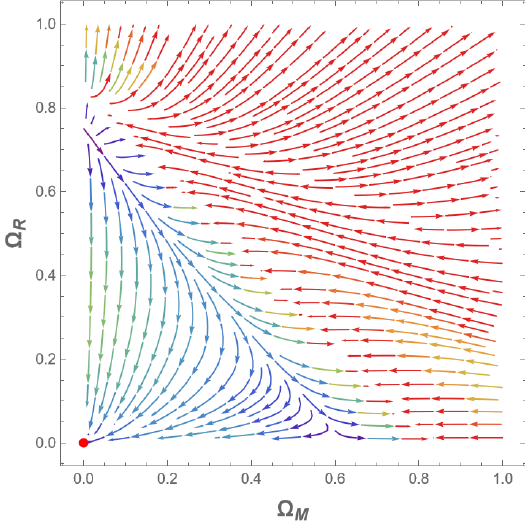}
\end{subfigure}
\hfill
\begin{subfigure}{0.45\textwidth}
\centering
\includegraphics[width=\linewidth]{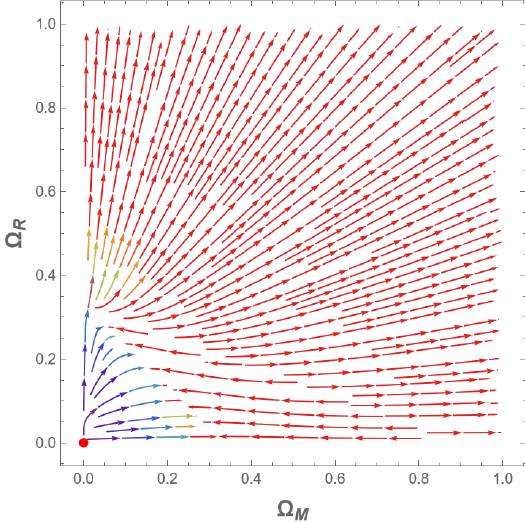}
\end{subfigure}
\caption{Local dynamics in the proximity of the point $P_1$ (red dot) for $k=1$ (left) and $k=-2$ (right). The remaining parameters are fixed as $p=1$, $\phi_0=1$, $\lambda=1$, $W_0=1$. The colors of the arrows indicate different magnitudes of the derivatives in each point, growing from blue to red.}
\label{kposneg}
\end{figure*}

In  \autoref{kposneg} we display two cross-sections of the phase space in the proximity of the fixed point $P_1$, for $k=1,-2$. The phase portrait are drawn with $p=1$ in both cases, because it turns out that the value of the parameter $p$ do not qualitatively change the general morphology of the vector field. Let us now analyze the behavior of the scale factor $a(t)$ when the dynamical system is in the unstable equilibrium given by the fixed points just described. As in the exponential potential case we find $x=0$, so that the scalar field $\phi$ results to be a constant and $Z(t)=H(t)$. We reconstruct the modified Hubble parameter from the value of the parameter $q$ in the fixed point, which in this case reads 
\begin{equation}
    q=-\frac{1}{k+1}
\end{equation}
from which
\begin{equation}
    Z(t)=\frac{1}{\frac{k}{k+1}t+c_1} \qquad c_1 \in \mathbb{R}.
\end{equation}
It is now immediate to evaluate the scale factor, resulting in
\begin{equation}
    a(t)= C_2\leri{\frac{k}{k+1}t+c_1}^{\frac{k+1}{k}} \qquad C_2>0.
\end{equation}
For $k$ taking integer values in the allowed set (which we remind to be $k\neq 0,-1$)
 the exponent $\frac{k+1}{k}$ is always positive, therefore the fixed points correspond to a spacetime expansion. By calculating the effective barotropic index, namely
\begin{equation}
    w_{\text{eff}}=-\frac{k+3}{3\leri{k+1}},
\end{equation}
we observe that for $k>0$ 
we describe accelerated expansions, with $-1<w_{\text{eff}}<-\frac{1}{3}$. For $k=-2$ and $k=-3$ we obtain $w_{\text{eff}}=\frac{1}{3}$ and $w_{\text{eff}}=0$ respectively, i.e. the spacetime dynamics is identical to radiation-dominated and matter-dominated expansions. Last, for $k\leq-4$ the fixed point equilibrium state describes other decelerated expansions, with $-\frac{1}{3}<w_{\text{eff}}\leq -\frac{1}{9}$. We point out that these fixed points correspond always to vacuum solutions ($\Omega_{M,R}=0$), where radiation and matter contributions are mimicked by the scalar field $\xi$ via a convenient choice of the potential parameters. It must be remarked that, contrary to the exponential potential case, here $u$ is not constrained to be positive. This implies that the phase space is no longer constrained to be be limited to values of the dynamical variables such that $u>0$ holds. However, an ulterior condition can be imposed on the state variables, namely $w_{\text{eff}} \leq 1$ corresponding to the request of having a sound speed in the effective cosmological fluid at most equal to the speed of light. This constraint allows to deal with a compact phase space, so that the analysis of the dynamics in the infinite regime can be disregarded. To conclude this section we summarize the results obtained in \autoref{tab: fixed points two fluid}.
\begin{table}[hbtp!]
\captionsetup{justification=justified, singlelinecheck=false}
\centering
\begin{tabular}[t]{V{1} C{1.8cm} | C{1cm} | C{5cm} |C{2cm}|C{1.7cm}|C{1.3cm}V{1}}
\hlineB{3}

\textbf{Potential}&\textbf{Point}& $\mathbf{y,\Omega_M,\Omega_R,u_\xi}$& \textbf{Existence} &\textbf{Stability}&$\mathbf{w_{\text{eff}}}$ \\[5pt]
\hlineB{2}

\multirow{8}{*}{E}
    &\multirow{2}{*}{$P_1$} & \multirow{2}{*}{$0,0,0,0$}& $s\le -1$ &S &   \multirow{2}{*}{$-1$} \\
    & & &  $s >-1$& A&  \\[5pt] \cline{2-6}

    &\multirow{2}{*}{$P_{2,3}$}& \multirow{2}{*}{$\pm\frac{\sqrt{2\leri{s+1}}}{\lambda},\,0,\,0,\,\pm\frac{\leri{s-2}\sqrt{2\leri{s+1}}}{\lambda}$}&  $-1 \leq s < 2$ & \multirow{2}{*}{S} &\multirow{2}{*}{$\frac{2s-1}{3}$} \\ 
    & & &$s\neq 0$& & \\[5pt] \cline{2-6}
 
    &$P_{4,5}$ & $\pm\frac{2}{\sqrt{s\lambda^2}},\,0,\,1-\frac{1}{s},\,\mp\frac{2}{\sqrt{s\lambda^2}}$ & $s\geq 1$ & S&  $\frac{1}{3}$ \\[5pt] \cline{2-6}
    &$P_{6,7}$  & $\pm\sqrt{\frac{3}{2s\lambda^2}},\,1-\frac{1}{2s},\,0,\,\mp\frac{3\sqrt{3}}{2\sqrt{2s\lambda^2}}$ &$s \geq \frac{1}{2}$& S&  $0$ \\[5pt]\hlineB{2}
 
 \multirow{2}{*}{P}
    &\multirow{2}{*}{$P_{1,2}$}  & \multirow{2}{*}{$\pm\sqrt{\frac{2k}{\lambda^2 \leri{k+1}}},0,0,\mp\sqrt{\frac{2k}{\lambda^2}}\frac{2k+3}{\leri{k+1}^{\frac{3}{2}}}$} & $p\le1$    &\multirow{2}{*}{S} &  \multirow{2}{*}{$-\frac{k+3}{3\leri{k+1}}$}\\
    & & &$k\neq-1,0$& & \\[5pt] \hlineB{3}
\end{tabular}
\caption{Properties of the fixed points in the presence of a cosmological fluid with a fixed radiation and matter component. For the sake of brevity, exponential and power-law potentials are denoted by capital E and P, respectively. In particular, in evaluating the fixed points for the exponential case we set $b=s\lambda^2$. For the stability properties we adopted the shortcut notation A, R and S, denoting respectively attractor, repeller and saddle points.}
 \label{tab: fixed points two fluid}
\end{table}

\section{General $w$ analysis}\label{sec: 6}
\begin{figure*}[htbp!]
\captionsetup{justification=justified, singlelinecheck=false}
\centering
\begin{subfigure}{0.45\textwidth}
\centering
\includegraphics[width=\linewidth]{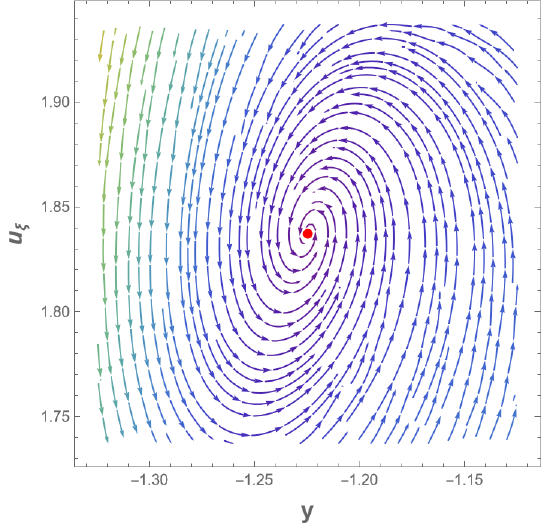}
\end{subfigure}
\hfill
\begin{subfigure}{0.45\textwidth}
\centering
\includegraphics[width=\linewidth]{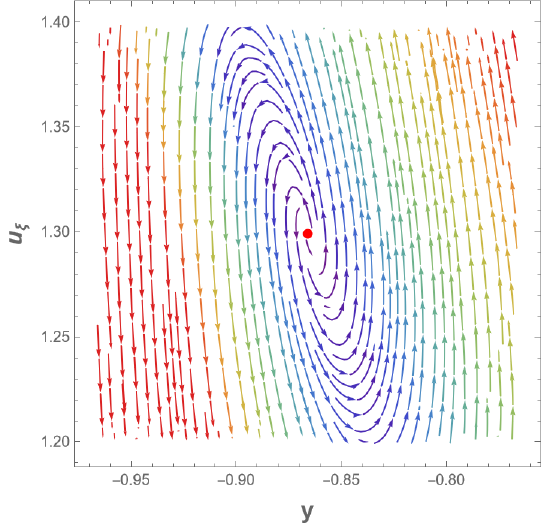}
\end{subfigure}

\vspace{0.5cm}

\begin{subfigure}{0.45\textwidth}
\centering
\includegraphics[width=\linewidth]{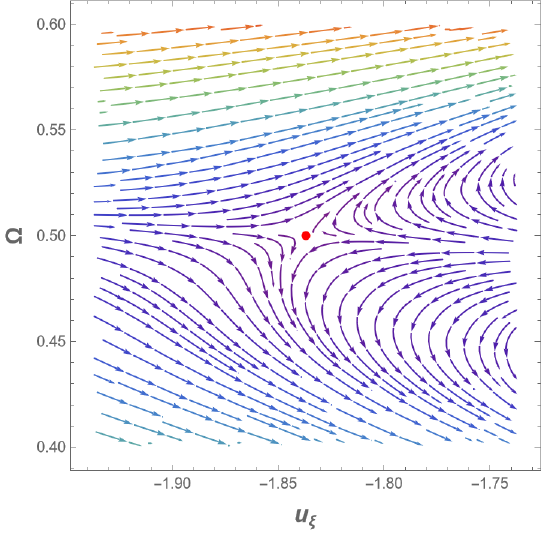}
\end{subfigure}
\hfill
\begin{subfigure}{0.45\textwidth}
\centering
\includegraphics[width=\linewidth]{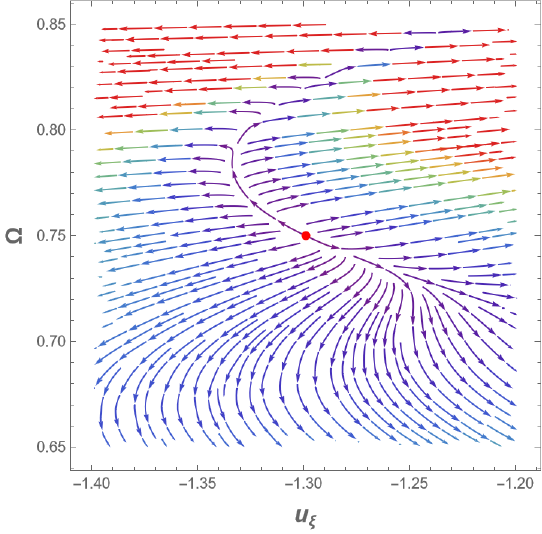}
\end{subfigure}
\caption{Transition of the point $P_4$ (red dot) from saddle (left) to repeller (right). The change of character is observed by varying the parameter $s$ from the value $s=1$ (left) to $s=2$ (right). The remaining parameters are set as $a=1$, $\lambda=1$, $W_0=1$, $w=0$.  }
\label{trans}
\end{figure*}

In this section we analyze the local behavior of the dynamical system when a single fluid, with barotropic index $w$, is assumed to be the only contribution in the energy momentum tensor. By considering the case of an exponential potential 
\begin{equation}\label{exppot}
W=W_0 e^{a \phi}\, e^{b \xi^2},
\end{equation}
we evaluate the fixed points admitted by the system, i.e.
\begin{align}
   &P_1=\leri{0,0,0} \\
   &P_{2,3}=\leri{\pm \frac{\sqrt{2\leri{s+1}}}{\lambda},0,\mp \frac{\leri{2-s}\sqrt{2\leri{s+1}}}{\lambda}}\\
    &P_{4,5}=\bigg( \pm \sqrt{\frac{3\leri{w+1}\leri{3w+1}}{2s\lambda^2}},1-\frac{\leri{3w+1}}{2s}, \mp\frac{3\leri{1-w}}{2}\sqrt{\frac{3\leri{w+1}\leri{3w+1}}{2s\lambda^2}} \bigg ),
\end{align}
where we have introduced, as in the previous section, the parameterization $b=s\lambda^2$. We select intervals of existence of such fixed points by requiring that $u$ and $\phi$ are positive and $y$ and $u_\xi$ real-valued, excluding in all cases the value $s=0$ because it implies a divergent scalar field $\xi$, as explained in the previous section. This conditions translate into the following set of inequalities for the models parameters: 
\begin{itemize}
\item The point $P_1$ exists for $a>0$.
\item Also points $P_{2,3}$ exist only for positive $a$, together with the supplemental condition $-1 \leq s <2$.
\item For the points $P_{4,5}$ existence is permitted also in the case of a negative $a$: in such case the ulterior conditions are $\frac{1}{3}<w<1$ and $s>\frac{\leri{3w+1}\leri{w+1}}{2\leri{3w-1}}$. On the contrary, by considering a positive $a$ we obtain: for $-1 \leq w < -\frac{1}{3}$ it must be $s \leq \frac{3w+1}{2}$; when instead  $ -\frac{1}{3}<w\leq \frac{1}{3}$ existence is guaranteed for $s \geq \frac{3w+1}{2}$; last, for $\frac{1}{3}<w<1$ we have $\frac{3w+1}{2} \leq s <\frac{\leri{3w+1}\leri{w+1}}{2\leri{3w-1}}$. We remark that the $w=-\frac{1}{3}$ case must be discarded because it implies a null potential. 
\end{itemize}
Let us now discuss the stability and the physics associated to each fixed point. 
\begin{itemize}
\item The eigenvalues of the Jacobian matrix evaluated at the origin of our system of coordinates, namely the fixed point $P_{1}$, result  
\begin{align}
    &a_1=-3 \leri{w+1}\\
    &a_{2,3}= \frac{-5 \pm \sqrt{1-24 s}}{2}.
\end{align}
Hence, for $w>-1$ we have that the point $P_1$ is, as in the two-fluids case, an attractor for $s>-1$ and a saddle if $s\leq -1$.
Instead, for $w<-1$  the fixed point result to be a saddle for every value of $s$. The particular case $w=-1$ needs a special attention: indeed, in this case we always have a null eigenvalue so that, while for $s<-1$ the points are saddles, being $a_{2,3}$ of different sign, for $s>-1$ the presence of two negative eigenvalues does not necessarily correspond to an attractive character of the fixed point and the dynamics in the proximity of the latter must be analyzed by resorting to center manifold techniques. Moreover, when $w=-1$ and $s=-1$ the point $P_1$ is characterized by two null eigenvalues, hence in this particular case we have to deal with a 2D center manifold. 
\item The fixed points $P_{2,3}$ result to be always saddles.
\item 
Regarding the stability $P_{4,5}$, we consider first the case of negative $a$, in which $P_{4,5}$ are the only fixed points of admitted by the dynamical system. It turns out that, when $s$ is chosen inside its allowed range, namely $s>\frac{\leri{3w+1}\leri{w+1}}{2\leri{3w-1}}$, these points are always saddles. Let us now turn to the case of positive $a$. Also for these points the case $w=-1$ presents some technical issues, given that the eigenvalues are $0,0,-5$ and a 2D center manifold analysis must be enforced. In the range $-1<w<-\frac{1}{3}$ $P_{4,5}$ are saddles. For $-\frac{1}{3}<w<\frac{1}{3}$ there is a number $s_0$, growing with $w$, such that $P_{4,5}$ are saddles  when $s<s_0$, whereas the latters become repellors for $s>s_0$. The specific value of $s_0$ can be numerically evaluated for any given value of $w$ in the range of interest. For $w=\frac{1}{3}$ the points $P_{4,5}$ are saddles. Last, in the interval $\frac{1}{3}<w<1$ we observe a situation similar to the one found in $-\frac{1}{3}<w<\frac{1}{3}$, namely a threshold value $s_0$ for which a change of behavior of the dynamics in the proximity of the fixed point is expected. Hence, as before the points $P_{4,5}$ are saddles for $s<s_0$ and become repellors for $s>s_0$. The only difference from the previous case is that the value of $s_0$ decreases with increasing $w$. 
\end{itemize}
We display, in  \autoref{trans}, phase portraits depicting the local dynamics in the proximity of the point $P_4$. As previously explained, for positive $a$ and $-\frac{1}{3}<w<\frac{1}{3}$, we expect a change in the nature of the fixed point, from saddle to repeller, depending on the value of $s$. Here we set $w=0$, for which we numerically evaluate $s_0 \simeq 1.44$. Then, we draw phase portraits corresponding to $s=1$ and $s=2$, in order to highlight the expected transition in the local character of the fixed point.

After having presented the stability analysis, let us now turn to the study of the scale factor behavior in correspondence with the fixed points of the system. For what concerns the points $P_1$ and $P_{2,3}$ we refer to the analysis provided in  \autoref{2fluexp}, given that for these points the scale factor behavior is identical to the two-fluids scenario. The fixed points $P_{4,5}$ correspond to matter scaling solutions, i.e. having effective barotropic index equal to that of the fluid $w_{\text{EFF}}=w$. It is found that they reproduce either accelerated expansions for $-1<w<-\frac{1}{3}$ or decelerated ones for $-\frac{1}{3}<w<1$. For $w=-1$ a de Sitter phase is reproduced. The scale factor in correspondence of such points is expressed $a(t) \propto t^{\frac{2}{3\leri{w+1}}}$.

It is interesting to consider an alternative choice of the potential $W$ corresponding to the power-law model
\begin{equation}\label{potpl}
    W=W_0\leri{\phi-\phi_0}^p \leri{\xi-\xi_0}^k
\end{equation}
where, as in the previous sections, $p$ and $k$ are integers numbers, $\phi_0$ and $\xi_0$ are reals with $\phi_0>0$. The fixed point we found are 
\begin{align}
    &P_1=\leri{0,0,0}\\
    &P_{2,3}=\leri{\pm\sqrt{\frac{2k}{\lambda^2 \leri{k+1}}},0,0,\mp\sqrt{\frac{2k}{\lambda^2}}\frac{2k+3}{\leri{k+1}^{\frac{3}{2}}} } ,
\end{align}
and the existence of such points is ensured, as before, by the constraints $k\neq0,-1$ and $p\leq 1$. Let us now provide the stability analysis for the fixed points above.
\begin{itemize}
    \item The Jacobian eigenvalues calculated in correspondence of point $P_1$ are 
    \begin{align}
        &a_1=-3\leri{w+1}\\
        &a_2=-3\\
        &a_3=-2.
    \end{align}
    Thus the fixed point $P_1$ is an attractor for $w>-1$ and a saddle for $w<-1$. The case $w=-1$ will be treated in the next section.

    \item The eigenvalues for the points $P_{2,3}$ are
     \begin{align}
        &a_1=-\leri{3w+1+\frac{2}{k+1}}\\
        &a_2=-4\\
        &a_3=2+\frac{1}{k+1}.
    \end{align}
    As in the two-fluids configuration these points correspond to saddles, given that $a_2$ and $a_3$ always have opposite signs.
\end{itemize}
It is immediate to provide the analysis of the physics associated to each fixed point. Indeed, given the values $q=-1$ and $x=0$, the origin $P_1$ represents a de Sitter phase. For what concerns the points $P_{2,3}$, the scale factor behavior is identical to the one described in Sec. \ref{powerlaw} for the points $P_{1,2}$. The results of the dynamical system analysis in the single-fluid case are collected in \autoref{tab: fixed points one fluid exponential} and \autoref{tab: fixed points one fluid power} for the exponential and power-law potential, respectively.

\begin{table*}[hbtp!]
\captionsetup{justification=justified, singlelinecheck=false}
\centering
\begin{tabular}{ V{1}  C{1cm} | C{5cm} | C{2.5cm} | C{2.8cm} | C{1.2cm}V{1} }
 \hlineB{3}
\textbf{Point}& $\mathbf{y,\Omega,u_\xi}$& \textbf{Existence} &\textbf{Stability}&$\mathbf{w_{\text{eff}}}$ \\[5pt]
 \hlineB{2}

\multirow{6}{*}{$P_1$} & \multirow{6}{*}{$0,0,0$}& \multirow{6}{*}{$a>0$} & S  & \multirow{6}{*}{$-1$} \\
     &   &  & $w < -1,\;\forall\; s$ & \\[5pt]\cline{4-4}
     &   &  & S & \\
     &   &  & $w > -1,\,s\le-1$ & \\[5pt]\cline{4-4}
     &  &  & A & \\
     &  &  & $w > -1,\,s>-1$ & \\[5pt]
    \Xcline{1-5}{0.7pt}

\multirow{2}{*}{$P_{2,3}$} & \multirow{2}{*}{$\pm \frac{\sqrt{2\leri{s+1}}}{\lambda},\,0,\,\mp \frac{\leri{2-s}\sqrt{2\leri{s+1}}}{\lambda}$}& $a>0,$ & \multirow{2}{*}{S} & \multirow{2}{*}{$\frac{2s-1}{3}$} \\
     &  &  $-1\le s \le 2$& & \\[5pt] 
    \Xcline{1-5}{0.7pt}

\multirow{13}{*}{$P_{4,5}$} & & $a<0,$& \multirow{3}{*}{S}& \multirow{13}{*}{$w$}\\
      & & $\frac{1}{3} < w < 1,$ & & \\
      & & $s>s_2$ & & \\[5pt]\Xcline{3-4}{0.7pt}
      & & $a>0,$ & \multirow{3}{*}{S}  & \\
      & & $-1 < w < -\frac{1}{3},$ & &\\
      & $\pm\sqrt{\frac{3(w+1)(3w+1)}{2s\lambda^2}},$ & $s\le s_1$ & & \\ [5pt]\Xcline{3-4}{0.7pt}
      &  $1-\frac{3w+1}{2s},$ & $a>0,$ & S $(s<s_0)$ & \\[5pt]\cline{4-4}
      & $\mp\frac{3(1-w)}{2}\sqrt{\frac{3(w+1)(3w+1)}{2s\lambda^2}}$ & $-\frac{1}{3}< w < \frac{1}{3}, $ & S $(w=\frac{1}{3})$ & \\[5pt]\cline{4-4}
      & & $s\ge s_1$ & R $(s>s_0)$ & \\ [5pt]\Xcline{3-4}{0.7pt}
      & & $a>0,$ & \multirow{2}{*}{S $(s<s_0)$ } & \\
      & & $\frac{1}{3}< w < 1, $ &  & \\ 
      & & $s\in \leri{s_1, s_2} $ &  R $(s>s_0)$ & \\[5pt]\hlineB{3}

\end{tabular}
\caption{Properties of the fixed points in the presence of a cosmological fluid with generic $w$ for the exponential potential. In particular, in evaluating the fixed points for the exponential case we set $b=s\lambda^2$. For the stability properties we adopted the shortcut notation A, R and S, denoting respectively attractor, repeller and saddle points. Boundaries values for the parameter $s$ are defined as $s_1=\frac{3w-1}{2}$ and $s_2=\frac{(3w+1)(w+1)}{2(3w-1)}$.}
\label{tab: fixed points one fluid exponential}
\end{table*}

\begin{table*}[hbtp!]
\captionsetup{justification=justified, singlelinecheck=false}
\centering
\begin{tabular}{ V{1}  C{1cm} | C{5cm} | C{2.5cm} | C{2.8cm} | C{1.2cm}V{1} }
 \hlineB{3}
\textbf{Point}& $\mathbf{y,\Omega,u_\xi}$& \textbf{Existence} &\textbf{Stability}&$\mathbf{w_{\text{eff}}}$ \\[5pt]
 \hlineB{2}

\multirow{2}{*}{$P_1$} &  \multirow{2}{*}{$0,0,0$} & $k\neq -1,0$& A $(w>-1)$ & \multirow{2}{*}{$-1$} \\[5pt]\cline{4-4}
 & & $p\le 1$ &S $(w<-1)$ & \\[5pt] \Xcline{1-5}{0.7pt}
 \multirow{2}{*}{$P_{2,3}$} & \multirow{2}{*}{$\pm\sqrt{\frac{2k}{\lambda^2 \leri{k+1}}},\,0,\,\mp\sqrt{\frac{2k}{\lambda^2}}\frac{2k+3}{\leri{k+1}^{\frac{3}{2}}}$} & $k\neq -1,0$ & \multirow{2}{*}{S} & \multirow{2}{*}{$-\frac{k+3}{3(k+1)}$} \\[5pt] 
  & & $p\le 1$ & & \\[5pt]
\hlineB{3}

\end{tabular}
\caption{Properties of the fixed points in the presence of a cosmological fluid with generic $w$ for the power-law potential. In particular, in evaluating the fixed points for the exponential case we set $b=s\lambda^2$. For the stability properties we adopted the shortcut notation A, R and S, denoting respectively attractor, repeller and saddle points. Boundaries values for the parameter $s$ are defined as $s_1=\frac{3w-1}{2}$ and $s_2=\frac{(3w+1)(w+1)}{2(3w-1)}$.}
\label{tab: fixed points one fluid power}
\end{table*}

\subsection{The cosmological constant scenario: $w=-1$.}
Here we discuss the particular case $w=-1$ which, as anticipated in the previous section, presents some technical issues due to the possible presence of a two-dimensional center manifold. Let use first enumerate the fixed points admitted by the system when the exponential potential \eqref{exppot} is assumed: they result to be
\begin{align}
   &P_1=\leri{0,\Omega,0} \\
   &P_{2,3}=\leri{\pm \frac{\sqrt{2\leri{s+1}}}{\lambda},0,\mp \frac{\leri{2-s}\sqrt{2\leri{s+1}}}{\lambda}}.
\end{align}
As opposed to the general $w$ case, here we have that for null $y$ and $u_\xi$ infinite fixed points are found, corresponding to a free choice of the value of the variable $\Omega$ in its range of existence, i.e. $0\leq\Omega\leq1$. In addition to this, we notice that the points $P_{4,5}$ described in the previous section are no longer fixed points for the system when the choice $w=-1$ is made. Let us now turn to the discussion of the existence of such points. 
\begin{itemize}
    \item The infinite set of fixed points, here concisely dubbed $P_1$, exists for $a>0$.
    \item The couple $P_{2,3}$ exist for $a>0$ and $-1\leq s\leq2$.
\end{itemize}
In order to assess the stability of the fixed points admitted by the system, we display the Jacobian eigenvalues for each equilibrium point.
\begin{itemize}
    \item Let us start by analyzing the local dynamics in correspondence of the fixed points $P_1$, for which the Jacobian eigenvalues read  
    \begin{align}
    &a_1=0\\
    &a_{2,3}= \frac{-5 \pm \sqrt{1-24 s\leri{1-\Omega}}}{2}.
\end{align}
When $s<-\frac{1}{1-\Omega}$ the points $P_1$ are saddles, given that $a_2$ and $a_3$ have opposite signs. For $s=-\frac{1}{1-\Omega}$ we have that both $a_1$ and $a_2$ are null and a center manifold analysis is required. For $s>-\frac{1}{1-\Omega}$ both $a_2$ and $a_3$ are negative but, in this case, a center manifold analysis is not necessary to assess the stability of the points. Indeed, in such a case it is found that the dynamics in the proximity of a fixed point $\leri{0,\Omega,0}$ is planar, namely the coordinate $\Omega$ is constant along the motion. Therefore, given the sign of $a_2$ and $a_3$ we claim that, within the interval $s>-\frac{1}{1-\Omega}$, the segment $\leri{0,\Omega,0}$, in which $0\leq \Omega\leq 1$, is a set of infinite attractors for the dynamical system. 
\item For what concerns the points $P_{2,3}$, the Jacobian eigenvalues are found to be 
\begin{align}
    &a_1=2\leri{s+1}\\
    &a_{2,3}=\frac{3s-2\pm\sqrt{s\leri{4-7s}+36}}{2}.
\end{align}
Calculations show that for $s>-1$ two eigenvalues are positive and one is negative, implying that $P_{4,5}$ are saddles. Instead, for $s=-1$, both points $P_{2,3}$ collapse into the origin. In this case, two eigenvalues result null, but we will provide the analysis of the local dynamics together with that of the point $P_1$ for $s=-1$ and $\Omega=0$.
\end{itemize}
 At this point, we set $s=-\frac{1}{1-\bar{\Omega}}$, with $0\leq\bar{\Omega}\leq 1$ a fixed value of the matter variable. In this setting two eigenvalues related to the local dynamics in the proximity of $P_1$ are null. We write the system in a set of coordinates in which the first order dynamics is diagonal, namely
 \begin{align}
     &q_1=\Omega-\bar{\Omega} \\
     &q_2=\frac{6y+2 u_\xi }{5}\\
     &q_3=\frac{-6y+3 u_\xi}{5}.
 \end{align}
 Having adopted this set of coordinates, the dynamical system boils down to
\begin{align}
    &q_1'=g_1(q_1,q_2,q_3)\\
    &q_2'=-5q_2+f(q_1,q_2,q_3)\\
    &q_3'=g_2(q_1,q_2,q_3).
\end{align}
We assume for the center manifold a second order Taylor expansion of the form 
\begin{equation}
    h\leri{q_1,q_3}=a_2 q_1^2+b_2 q_3^2+c_2q_1q_3
\end{equation}
and we solve the determinant equation \eqref{detcenter} in order to find the resolving values for the coefficients $a_2$, $b_2$ and $c_2$. Explicit calculations show that the center manifold is written at second order as
\begin{equation}
    h\leri{q_1,q_3}=\frac{4s}{25}q_1q_3
\end{equation}
and the dynamics restricted to such center manifold is determined by the dynamical system
\begin{align}
    &q_1'=\frac{\lambda^2}{9}\leri{1+\frac{1}{s}}q_3^2\\
    &q_3'=\frac{6s}{5}q_1 q_3.
\end{align}
The system above admits an exact solution given by
\begin{align}
    &q_1(t)=\sqrt{\frac{5\lambda^2c_1 |s+1|}{27s^2}} \tanh{\leri{\sqrt{\frac{\lambda^2 c_1 |s+1|}{15}} \leri{2t+45c_2 s}}} \\
    &q_3(t)=-\sqrt{\frac{4c_1}{1+\cosh \leri{\sqrt{\frac{4\lambda^2 c_1 |s+1|}{15^3}}\leri{2t+45c_2 s}}}}   
\end{align}
where $c_1>0$ and $c_2\in \mathbb{R}$ are integration constants. We observe that the trajectory remains finite in the limit $t \to \infty$, where $\Omega=q_1+\bar{\Omega}$ approaches the asymptotic value
\begin{equation}
    \Omega = \bar{\Omega}+ \sqrt{\bar{\Omega}^2+\frac{5\lambda^2c_1\bar{\Omega}(1-\bar{\Omega})}{27}}>\bar{\Omega},
\end{equation}
implying the stability of the local dynamics restricted to the center manifold. The special sub-case corresponding to $s=-1$, or equivalently $\bar{\Omega}=0$, is an example of 2D restricted dynamics in the $\leri{y,u_\xi}$ plane. As explained before, in such conditions the dynamical system admits a unique fixed point in the origin, with eigenvalues $a_1=0$ and $a_2=-5$. Following the same steps described in the previous sections, we determine the character of the local dynamics for the coordinate $q$ on the center manifold via the differential equation
\begin{equation}
q'=\frac{\lambda^2}{15}q^3 + \mathcal{O}\leri{q^5},
\end{equation}
 predicting instability along the direction given by $q$. Therefore we claim that, in the scenario $s=-1$ and $\bar{\Omega}=0$, the origin $\leri{y,\Omega,u_\xi}=\leri{0,0,0}$ is a saddle for the dynamical system.

 Let us now analyze the case of the dynamical system for $w=-1$ when a power-law potential as in \eqref{potpl} is chosen. The fixed points are 
 \begin{align}
   &P_1=\leri{0,\Omega,0} \\
   &P_{2,3}=\leri{\pm\sqrt{\frac{2k}{\lambda^2 \leri{k+1}}},0,0,\mp\sqrt{\frac{2k}{\lambda^2}}\frac{2k+3}{\leri{k+1}^{\frac{3}{2}}} }.
\end{align}
As in the case of an exponential potential, we calculate an infinite set of fixed points corresponding to the segment $0\leq \Omega \leq 1$. The existence of such points is related to the constraint $p<\frac{2}{1-\Omega}$. For what concerns $P_{2,3}$, their existence is ensured, as in the general $w$ case, for $k\neq0,1$ and $p\leq1$. The stability analysis follows from the calculation of the eigenvalues in correspondence to each fixed point, resulting in 
\begin{itemize}
    \item The eigenvalues for the point $P_1$ are
    \begin{align}
        &a_1=0\\
        &a_2=-3\\
        &a_3=-2.
    \end{align}
    As in the case of an exponential potential, here we can affirm that the dynamics in the proximity of each fixed point of the form $\leri{0,\Omega,0}$ is stable without resorting to center manifold analyses. Indeed, when a small perturbation is induced, the trajectory results planar and the coordinate $\Omega$ is constant along the motion. 

    \item For what concerns the points $P_{2,3}$ we calculate
 \begin{align}
        &a_1=\frac{2k}{k+1}\\
        &a_2=-4\\
        &a_3=2+\frac{1}{k+1}.
    \end{align}
    Given that at least two eigenvalues have different signs for every value of the parameter $k$ we can affirm that these points corresponds to saddles for the dynamical system.
\end{itemize}
As in the previous sections, we display our findings in the cosmological constant scenario in \autoref{tab: fixed points center manifold}.

\begin{table*}
\captionsetup{justification=justified, singlelinecheck=false}
\centering
\begin{tabular}{ V{1} C{1.8cm} | C{1cm} | C{5cm} | C{2cm}|C{1.7cm} | C{1.3cm}V{1} }
 \hlineB{3}
 \textbf{Potential}&\textbf{Point}& $\mathbf{y,\Omega,u_\xi}$& \textbf{Existence} &\textbf{Stability}&$\mathbf{w_{\text{eff}}}$ \\[5pt]
 \hlineB{2}

\multirow{8}{*}{E} & \multirow{6}{*}{$P_1$} & \multirow{6}{*}{$0,\,\Omega,\,0$} & \multirow{6}{*}{$a>0$} & S & \multirow{6}{*}{-1}\\[5pt]

 & & & & $s<\frac{1}{\Omega-1}$ & \\[5pt]\cline{5-5}

 & & & & A & \\[5pt]
 & & & & $s=\frac{1}{\Omega-1}$& \\[5pt]\cline{5-5}

 & & & & A & \\[5pt]
 & & & & $s>\frac{1}{\Omega-1}$& \\[5pt]\Xcline{2-6}{0.7pt}

& \multirow{2}{*}{$P_{2,3}$} & \multirow{2}{*}{$\pm \frac{\sqrt{2\leri{s+1}}}{\lambda},\,0,\,\mp \frac{\leri{2-s}\sqrt{2\leri{s+1}}}{\lambda}$} & $a>0,$ & S  & \multirow{2}{*}{$\frac{2s-1}{3}$} \\[5pt]

& & & $-1\le s\le2$ & $s>-1$ & \\[5pt] \hlineB{2}

\multirow{3}{*}{P} & $P_1$ & $0,\,\Omega,\,0$ & $p<\frac{2}{1-\Omega}$ & A & -1 \\[5pt]\Xcline{2-6}{0.7pt}

 & \multirow{2}{*}{$P_{2,3}$} & \multirow{2}{*}{$\pm\sqrt{\frac{2k}{\lambda^2 \leri{k+1}}},\,0,\,\mp\sqrt{\frac{2k}{\lambda^2}}\frac{2k+3}{\leri{k+1}^{\frac{3}{2}}}$} & $k\neq 0,1$ & S & \multirow{2}{*}{$-\frac{k+3}{3(k+1)}$}\\[5pt]
 
  & & & $p\le 1$ & & \\[5pt]\hlineB{3}
\end{tabular}
\caption{Properties of the fixed points for $w=-1$. For the sake of brevity, exponential and power-law potentials are denoted by capital E and P, respectively. In particular, in evaluating the fixed points for the exponential case we set $b=s\lambda^2$. For the stability properties we adopted the shortcut notation A, R and S, denoting respectively attractor, repeller and saddle points.}
\label{tab: fixed points center manifold}
\end{table*}
\newpage
\section{Reconstruction of the original $f(\pal,\xi,X)$ model}\label{sec: 7}
In this section we derive the explicit functional forms of the Lagrangian densities corresponding to the choices for the potential in \eqref{eq: potential forms}. We start by the general definition of the generalized potential $U$, i.e.
\begin{equation}
    U\leri{\frac{\partial f}{\partial \pal}, \xi, X}= \pal\frac{\partial f}{\partial \pal}-f\leri{\pal,\xi,X},
\end{equation}
where now we are assuming $\phi=\frac{\partial f}{\partial \pal}$. This represents a Clairaut's equation for the unknown function $f$, which upon a rearrangement of the terms and derivation w.r.t. $\pal$ can be put into the form:
\begin{equation}
    \frac{\partial^2 f}{\partial \pal^2}\leri{\pal - \left.\frac{\partial U}{\partial \phi}\right|_{\phi=f_\pal}}=0.
\end{equation}
The so-called general solution is simply obtained by requiring $f_{\pal\pal}=0$, resulting in the functional form
\begin{equation}
    f(\pal,\xi,X)=\alpha(\xi,X)\pal + \beta(\xi,X),
\end{equation}
where $\alpha, \beta$ are two undetermined function of the scalar field and its kinetic term. Here, we are interested instead in the singular solutions, which can be obtained by demanding
\begin{equation}
    \pal - \left.\frac{\partial U}{\partial \phi}\right|_{\phi=f_\pal}=0.
\end{equation}
For the exponential potential, this equation reduces to
\begin{equation}
    \pal -\lambda^2 X - a W_0 e^{b\xi^2}e^{a f_\pal}=0,
\end{equation}
which can be rewritten as
\begin{equation}
    \frac{\partial f}{\partial \pal} = \ln \leri{\frac{\pal-\lambda^2 X}{a W_0 e^{b\xi^2}}}^{\frac{1}{a}}.
\end{equation}
The solution can be obtained by direct integration and after some manipulation can be put into the form
\begin{equation}
    f(\pal,\xi,X)=f_0(\xi,X)+\frac{\pal-\lambda^2 X}{a}\ln\frac{\pal-\lambda^2 X}{a W_0 e ^{b\xi^2+1}},
    \label{eq: f form exp pot}
\end{equation}
where we defined the purely $k$-essential term $f_0(\xi,X)$ as
\begin{equation}
    f_0(\xi,X)\equiv f(\pal_0,\xi,X)-\frac{\pal_0-\lambda^2 X}{a}\ln\frac{\pal_0-\lambda^2 X}{a W_0 e ^{b\xi^2+1}},
\end{equation}
with $\pal_0$ some constant Palatini curvature reference value. For consistency, we also assumed $\pal-\lambda^2 X > 0$, with $a, W_0 >0$ already by hypothesis. We note that similar logarithmic corrections, at least in the context of ordinary $f(R)$ gravity, usually trace back to quantum gravity effects, as discussed in \cite{Olmo:2008nf,Sadeghi:2015nda,Odintsov:2017hbk,Delhom:2023xxp}.
\\ For what concerns the power law potential, the singular solution to the Clairaut's equation is determined in this case by
\begin{equation}
    \pal -\lambda^2 X - p W_0 \leri{\frac{\partial f}{\partial \pal}-\phi_0}^{p-1}\leri{\xi-\xi_0}^k = 0,
\end{equation}
which, provided $p\neq0,1$, can be put into the form
\begin{equation}
    \frac{\partial f}{\partial \pal} = \phi_0 + \leri{\frac{\pal -\lambda^2 X}{p W_0 \leri{\xi-\xi_0}^k}}^{\frac{1}{p-1}}.
\end{equation}
The solution can be obtained again by direct integration and up to a redefinition of terms it can be rearranged as
\begin{equation}
    f(\pal,\xi,X)=f_0(\xi,X)+\phi_0\pal+(p-1)W_0\leri{\xi-\xi_0}^{k}\leri{\frac{\pal-\lambda^2 X}{p W_0\leri{\xi-\xi_0}^{k}}}^{\frac{p}{p-1}},
    \label{eq: functional form f power pot}
\end{equation}
where now $f_0(\xi,X)$ is given by
\begin{equation}
    f_0(\xi,X) \equiv -\phi_0\pal_0+f(\pal_0,\xi,X)-(p-1)W_0\leri{\xi-\xi_0}^{k}\leri{\frac{\pal_0-\lambda^2 X}{p W_0\leri{\xi-\xi_0}^{k}}}^{\frac{p}{p-1}}.
\end{equation}
We recall from \autoref{sec: 5} and \autoref{sec: 6} that $p<0$ and $k\neq 0,1$ (with $p,k\in\mathbb{Z}$), so that when $p-1$ is even we have to require the additional condition
\begin{equation}
    \frac{\pal-\lambda^2 X}{ W_0\leri{\xi-\xi_0}^{k}}<0.
\end{equation}
The functional form \eqref{eq: functional form f power pot} can be seen as some sort of generalization of Born-Infeld-Einstein inspired models \cite{Vollick:2005gc,Burrage:2014uwa,Afonso:2017aci,Benisty:2021laq}, displaying en explicit coupling with $k$-essence terms.
\section{Discussion of the results and final remarks}\label{sec: 8}
In this work we formulated an extended $k$-essence model, where the scalar field and its kinetic term are non-minimally coupled to the gravitational sector, which we built in analogy with the Palatini $f(\pal)$ theory of gravity. More specifically, we considered for the Lagrangian density a function depending on the Ricci curvature $\pal$, the scalar field $\xi$ and the kinetic term $X=\nabla_\mu\xi\nabla^\mu\xi$. We tackled the model in its equivalent scalar-tensor reformulation, resulting in a pair of additional scalar fields, with the typical scalaron $\phi$ of the Jordan frame still playing the role of an auxiliary field, as for standard Palatini $f(\pal)$ theories. Concretely, we derived a generalized structural equation \cite{Olmo:2011uz}, allowing in principle to algebraically resolve for the scalaron in terms of $\xi$, its kinetic term and, possibly, the trace of the energy momentum tensor of additional matter. This implied the emergence in the equation for $\xi$ of novel derivative and interaction terms, together with non-canonical matter sources, with respect to the case of a $k$-essence field in General Relativity. In particular, the effective metric defining the principal part of the equation of motion for $\xi$ is modified, leading to the loss of a conformal relation with the original spacetime metric even for a Lagrangian density linearly depending on the kinetic term. This is due to the appearance in the expression for $Q^{\mu\nu}$ of an additional term in front of the disformal part, which is in general not vanishing even if $U$ is simply proportional to $X$, as it occurs instead for the original $k$-essence model~\cite{Babichev:2007dw}. The same contribution is also present in the condition defining the hyperbolicity of the wave operator for $\xi$, assuring the well-posedness of the initial value problem for the evolution of the scalar field. Concerning the absence of dynamical instabilities, results from standard $k$-essence theories were recovered, and we simply have to require the positiveness of $U_X$. To be exact, in our case such a constraint must be accompanied by the supplementary condition $\phi>0$, as it is usually assumed in Palatini $f(\pal)$ theories, in order to guarantee a positive effective gravitational coupling in the weak field limit\footnote{More in general, the condition $\phi>0$ is also demanded for consistency in the definition of the conformal metric in Einstein frame  \cite{Sotiriou2007}}. We remark that a simple Palatini $f(\pal)$ model, supplemented by a pure $k$-essence term $K(\xi,X)$, is actually a subclass of our general formulation, when from the generalized potential $U$ is singled out the dependence on $\phi$, namely $U(\phi,\xi,X)=V(\phi)-K(\xi,X)$. Here $V(\phi)$ can be indeed identified with the usual potential term for the scalar-tensor representation of Palatini $f(\pal)$ theories, with the result that in the structural equation the contribution of the different scalar fields is now disentangled (see \autoref{app: a}).
\\In vacuum, where it is possible to consider for $\phi$ the solution $\phi=\phi(\xi,X)$,  we also established a formal relation between the resulting effective action \eqref{eq: action f(R)} and the class Ia of DHOST theories \cite{BenAchour:2016cay}. Such a correspondence is explicitly broken in the presence of matter, with the introduction of a non-minimal coupling between the trace of the energy momentum tensor and the dynamical degrees of freedom. These properties suggest the possible relevance of our theoretical setting in addressing the DESI results \cite{DESI:2025fii,DESI:2025wyn,DESI:2025zgx}, in that it combines in a single framework the different strategies pursued for evading the No-go theorem associated to Quintom scenarios \cite{Cai:2009zp,Cai:2025mas}. Our Palatini $k$-essence model, is able indeed to include both higher-derivative terms for the additional scalar degree of freedom and to introduce a non-trivial coupling of the latter with the cosmological fluid, retaining the properties of the DHOST approach, as well as those of the Palatini formulation. As a preliminary step, in this work we chose to pursue a minimal prescription on the form of the function $U$, by requiring that the resulting structural equation does not contain any explicit dependence of $X$, but still involving the presence of external matter\footnote{We point out that in this case, in vacuum, the additional property $\phi_X=0$ gives us $\alpha_4=0$, spoiling the equivalence with the DHOST theories, since for this configuration all the functions $\alpha_i$ defined in \cite{BenAchour:2016cay,Langlois:2015cwa} are vanishing.}. This led us to consider the form $U=\lambda^2\phi X + W(\phi,\xi)$, for which the requirements of the hyperbolicity of the equation for $\xi$ and the absence of ghost instabilities boiled down to the same condition $U_X>0$, identically satisfied for the parametrization $\lambda^2$ adopted.
\\We then specialized the analysis to a FLRW cosmological setting. As a first step, we introduced the modified Hubble function $Z=H+\frac{\dot{\phi}}{2\phi}$, thanks to which we were able to absorb the second time derivatives of the scalar field $\phi$ and write the cosmological evolution equations as a set of first order ODEs, except for \eqref{eq: xi frw z}, giving the evolution of $\xi$, which remains of the second order. We then downgraded every first order ODE to a constraint equation, by means of the set of dimensionless variables provided in \eqref{sdv1}-\eqref{def uxi}. We also lowered the order of the equation for $\xi$ by writing it in terms of $y=\xi'$, i.e. the scalar field velocity in the dimensionless time $\tau=\ln \leri{\sqrt{\phi}a}$. In order to obtain an autonomous system, we derived two other dynamical equations for the dimensionless matter variable $\Omega$ and the gradient of the potential term $u_\xi$. It must be remarked that, in this work, we have restricted our attention to potential functions of the form $W(\phi,\xi)=W_0 g(\phi) f(\xi)$ in which the dependence from each scalar field can be factorized. For this class of potentials, the inversion formulae given in \eqref{eq: inv g} and \eqref{eq: inv f} permit to calculate every quantity of the constrained set $\leri{q,x,u,u_\phi}$ in terms of the dynamical variables $\leri{y,\Omega, u_\xi}$. However, we stress that the choice of a factorizable potential is not mandatory in order to achieve an autonomous dynamical system describing the cosmological evolution. Indeed, as already stated in \autoref{sec: 4}, more general forms are feasible for the potential $W$, as long as the relations \eqref{eq: inv g}-\eqref{eq: inv f} can be solved for the scalar fields $\phi$ and $\xi$.\\
For the sake of simplicity, we considered only two concrete examples for the functions $g(\phi)$ and $f(\xi)$, namely the exponential and the power-law forms. For what concerns the matter content, we have assumed first a cosmological fluid composed by matter and radiation and, as a second part, we have examined the case of a single matter component, but with unspecified equation of state parameter $w$. In both these scenarios we can outline the presence of common traits, signaling robust features of the cosmological evolution:
\begin{itemize}
    \item for both the exponential and power-law potential, the origin results to be either an attractor or a saddle. This equilibrium point is always associated with a de Sitter phase: the reason for this is that $\xi$ is constant in this configuration and the associated potential term plays the role of a cosmological constant, implying an exponential profile of the scale factor;
    \item with great generality we observe the presence of equilibrium points corresponding to vacuum solutions, in which the dynamics is dominated by the kinetic and potential energy of the scalar field $\xi$. These equilibria correspond to points $P_{2,3}$ in all cases but the power-law potential for the two fluids configuration, in which are instead dubbed $P_{1,2}$. The effective barotropic index here is a function of the theory parameters and the physics associated can therefore mimic a plethora of diverse phenomenologies; 
    \item the exponential potential admits, in specific regions of the parameters space, the existence of scaling solutions, where the effective barotropic index exactly matches the bare barotropic index of the dominant fluid component. These solutions always correspond to saddles.
\end{itemize}
To summarize, the Palatini $k$-essence model here investigated shows a rich phenomenology when applied to cosmology. The presence of scaling solutions, which we have shown to be saddles for the dynamical system, offer the possibility for prolonged, but transient, phases dominated by one of the cosmological fluid components, reproducing the observed matter and radiation eras. The ubiquitous presence of the de Sitter attractor in the origin signals the feasibility for both an inflationary and a late-time exponential expansion. Moreover, the possibility for this attractor to be connected to other equilibrium points by heteroclinic orbits implies that the dynamical system is very likely to naturally evolve towards a final de Sitter phase. For what concerns the case of a single component cosmological fluid we outline that here scaling solutions are allowed to behave as repellors for certain values of the theory and physical parameters. This feature can be considered as a potential tool to discriminate between physically reasonable regions of the model, if one elects the presence of a fluid-dominated phase as a viability criterion to constraint the theory. In this respect, the analysis of the two-fluid configuration has shown that the power-law potential must be regarded as strongly disfavored by observations. Indeed, when both matter and radiation are included ab initio in the energy momentum tensor, a viability criterion for a specific choice of the potential must require the appearance of equilibrium points describing matter and radiation dominated phases for the dynamical system evolution and, as displayed in \autoref{tab: fixed points two fluid}, this property is not satisfied by the power-law potential.
\\The special sub-case of a cosmological constant, reproduced by setting $w=-1$, is characterized by the presence of a continuous line of infinite equilibrium points of the form $\leri{0,\Omega,0}$. When the system is close to one of such points, the dynamics becomes essentially planar and the evolution takes place at constant $\Omega$. Therefore, the assumption of a cosmological constant component in the matter sector introduces a possible degeneracy in the system evolution. Indeed, a de Sitter expansion can be associated either with a fluid or a scalar field dominated phase, as for instance when the origin is considered.
Summing up, the most promising scenario, able to reconstruct a coherent sequence of radiation and matter dominated phases, ending in a late-time De Sitter expansion where a constant value of the scalar field mimics the effect of a cosmological constant, follows from the choice of the exponential potential. Moreover, this configuration is actually robust, in the sense that for its existence a fine-tuning of the parameters is not required. However, we have to remark that it is not possible to reproduce a phantom-dominated phase for this system, nor a crossing from below of the $-1$ line for the parameter $w_{\text{eff}}$, as suggested by recent DESI observations. This does not rule out the general capability of this model to give account of such phenomenology, but rather could indicate a limitation of the simple sub-case here considered. For instance, by considering $m\neq1$ in \eqref{formau} or more complicated, non-factorizable forms of the potential (still preserving the invertibility property necessary to the well-posedness of the dynamical system), an even richer phenomenology is expected to result.
\\ Testing the actual robustness of the model, in reproducing consistent dynamical phases of the evolution of the Universe, requires the implementation in the analysis of numerical techniques, in order to deal with more complex form for the generalized potential $U(\phi,\xi,X)$. Relaxing the conditions of an explicit algebraic resolvability for the field $\phi$ in the structural equation, or the feasibility of the inversion in \eqref{eq: inv g}, \eqref{eq: inv f}, allows us indeed to consider a wider set of couplings between the auxiliary field $\phi$ and the k-essence scalar field $\xi$. These generalizations are currently under investigations and the results will be published in a forthcoming work. An alternative route, still relying on a dynamical system analysis, is instead represented by the formalism developed in \cite{Jarv:2021qpp,Jarv:2024krk}, which turned out to be capable of assessing some "model independent" conditions guaranteeing the emerging of well-posed inflationary scheme. These, in particular, also concern the properties of perturbations. By introducing modifications in the FLRW metric, representing inhomogeneities and anisotropies, it could be possible to determine the stability of the identified fixed points via linear perturbation theory. This should yield testable predictions to be compared with the Cosmic Microwave Background anisotropy and the large-scale structure of the Universe. 
\\ We conclude the discussion by noting that a possible future analysis of the weak-field limit of the theory can allow to constraint the theory parameter space. Concretely, the model could be tested in order to verify its viability with respect to local Solar System physics and gravitational waves morphology to be compared with current observations.

\begin{acknowledgments}
The work of FB has been supported by grants PID2020-116567GB-C21, PID2023-149560NBC21 funded by MCIU/AEI/10.13039/501100011033 and FEDER, UE. This article is based upon work from COST Action CA21136 Addressing observational tensions in cosmology with systematics and fundamental physics (CosmoVerse) supported by COST (European Cooperation in Science and Technology). Authors thank Renello B. Magalh\~aes for his insightful suggestions about the visualization of the results.
\end{acknowledgments}

\appendix{}
\section{Palatini $f(\pal)$ gravity with a $k$-essence field}\label{app: a}
The aim of this appendix is to elucidate the differences between our model and a Palatini $f(\pal)$ theory in the presence of a $k$-essence scalar field. We start by noticing that at the level of the scalar-tensor representation the second case can be formally obtained by simply setting
\begin{equation}
    U(\phi,\xi,X)=V(\phi)-K(\xi,X).
\end{equation}
This implies that the structural equation reduces to
\begin{equation}
    2V(\phi)-\phi V'(\phi)-2K(\xi,X)+XK_X(\xi,X)=\kappa^2 T,
\end{equation}
so that the auxiliary field $\phi$ can be still formally solved in terms of $\xi$ and its kinetic term. In particular, in this case demanding for $\phi$ to be independent of $X$, leads to the condition $K\propto g(\xi)X^2$, which actually implies also $\phi=\phi(T)$. The equation for the scalar field is given by
\begin{equation}
\Box\xi-\frac{K_\xi}{2K_X}+\frac{\nabla^\mu\xi}{K_X}\leri{K_{X\xi}\nabla_\mu\xi+K_{XX}\nabla_\mu X}=0
\label{eq: scalar xi palatini pure},
\end{equation}
where every dependence on $\phi$ has now dropped out. Therefore, the dynamics of $\xi$ does not receive any contribution from the structural equation, and no source terms depending on the trace of the energy momentum tensor are introduced. Then, taking into account that now $U_X = - K_X$, the hyperbolicity of the equation of motion is guaranteed for 
\begin{equation}
    2X K_{XX}+K_X >0,
\end{equation}
in agreement with the standard result of \cite{Babichev:2007dw,Babichev:2016hys}.
\section{Eigenvalues of the Jacobian matrix}\label{app:b}
In this appendix we report for the sake of completeness the expression of the eigenvalues of the Jacobian matrix, evaluated at the fixed points, for the exponential potential when two cosmological fluids are considered (see discussion in  \autoref{sec: 5}). These are given by, respectively:
\begin{itemize}
    \item Point $P_1$: $a_1=-4,\;a_2=-3,\;a_{3,4}=\frac{ -5 \pm \sqrt{1 - 24 s}}{2}$
    
    \item Points $P_{2,3}$: $a_1= 2 s-1,\;a_2=2 (s-1),\;a_3=\frac{4 + 
 s ( 3 s-8)}{2(s-2)}-\Lambda_1(s),\;a_4=-1 + \frac{3 s}{2} + \Lambda_1(s)$, 
\\with $\Lambda_1(s)=\frac{\sqrt{( s+2) ( 18-7 s)}}{2}$
    \item Points $P_{4,5}$: $a_1=1,\;a_2=\frac{1}{3}-\frac{\Sigma_1(s)}{\Sigma_2(s)}+\frac{\Sigma_2(s)}{3s^4},\;a_{4,5}=\frac{1}{3}+\frac{1}{2}\leri{\frac{\Sigma_1(s)}{\Sigma_2(s)}-\frac{\Sigma_2(s)}{3s^4}}\pm i\sqrt{3}\leri{\frac{\Sigma_1(s)}{\Sigma_2(s)}+\frac{\Sigma_2(s)}{3s^4}}$,
    \\with $\Sigma_1(s)=12 s^3 - \frac{11}{3} s^4,\; \Sigma_2(s)=(s^{11} ( 199 s-162) + 
        6 \sqrt{3} \sqrt{ \Sigma_3(s) })^\frac{1}{3}$ and \\$\Sigma_3(s)=s^{21}\leri{379 s^3-718 s^2+ 639 s-432}$

    \item Points $P_{6,7}$: $a_1=-1,\;a_2=\frac{\leri{\Theta_1(s)+\frac{s^2 \Theta_2(s)^{3/2} \Theta_3(s)}{\Theta_6(s)}+\Theta_2(s)^{9/2}\Theta_6(s)^3}^{1/3}}{6s^{3/2}\Theta_2(s)^2}$ and \\$a_{4,5}=-\frac{1}{6}-\frac{(1\pm i\sqrt{3})\sqrt{s}\Theta_3(s)}{12 \sqrt{\Theta_2(s)}\Theta_6(s)}-\frac{(1\mp i\sqrt{3})\Theta_6(s)}{s^\frac{3}{2} \sqrt{\Theta_2(s)}}$, where we defined    
\begin{align}
    \Theta_1(s) & = -9 s^\frac{3}{2} + 12 s^\frac{5}{2} - 52 s^\frac{7}{2} + 32 s^\frac{9}{2} - 64 s^\frac{11}{2}\nonumber\\
    \Theta_2(s) & =8s^2-2s+3\nonumber \\
    \Theta_3(s)&=8 s^3- 2 s^2+ 111s +162\nonumber \\
    \Theta_4(s) & =-s^6(2048 s^8- 498688 s^7+ 
        167040 s^6 + 114208 s^5- 71912 s^4 +\nonumber\\ 
        &\;\;\;\;\;+78396 s^3 + 31194 s^2 + 17739 s +17496)\nonumber\\
    \Theta_5(s) & =8 s^3- 3890 s^2+ 165s+1215 \nonumber\\
    \Theta_6(s) & = 9\sqrt{3 \Theta_4(s)}-s^\frac{7}{2} \sqrt{\Theta_2(s)} \Theta_5(s) \nonumber\\  
\end{align}
\end{itemize}
\begin{figure*}[h]
\captionsetup{justification=justified, singlelinecheck=false}
\centering
\includegraphics[width=7.3cm]{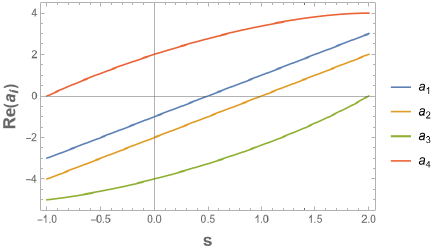}\hspace{0.3 cm}\includegraphics[width=7.3cm]{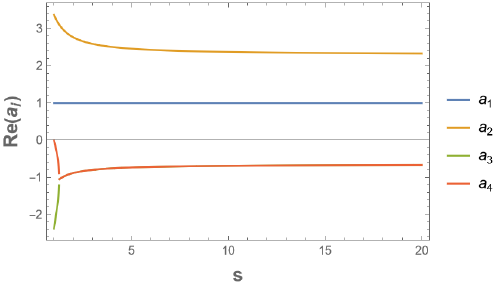}
\caption{Real part of the Jacobian eigenvalues calculated for the points $P_{2,3}$ (left) and $P_{4,5}$ (right).}
\label{autovaloridoppioexp}
\end{figure*}

\bibliographystyle{JHEP}
\bibliography{references}
\end{document}